# Discovery of peculiar periodic spectral modulations in a small fraction of solar type stars

Ermanno F. Borra and Eric Trottier

Département de Physique, Université Laval, Québec, Qc, Canada G1V 0A6

(email: borra@phy.ulaval.ca)

**KEY WORDS:** Stars: solar type - Techniques: spectroscopic - Extraterrestrial intelligence

**ABSTRACT**

A Fourier transform analysis of 2.5 million spectra in the Sloan Digital Sky Survey was carried out to detect periodic spectral modulations. Signals having the same period were found in only 234 stars overwhelmingly in the F2 to K1 spectral range. The signals cannot be caused by instrumental or data analysis effects because they are present in only a very small fraction of stars within a narrow spectral range and because signal to noise ratio considerations predict that the signal should mostly be detected in the brightest objects, while this is not the case. We consider several possibilities, such as rotational transitions in molecules, rapid pulsations, Fourier transform of spectral lines and signals generated by Extraterrestrial Intelligence (ETI). They cannot be generated by molecules or rapid pulsations. It is highly unlikely that they come from the Fourier transform of spectral lines because too many strong lines located at nearly periodic frequencies are needed. Finally we consider the possibility, predicted in a previous published paper, that the signals are caused by light pulses generated by Extraterrestrial Intelligence to makes us aware of their existence. We find that the detected signals have exactly the shape of an ETI signal predicted in the previous publication and are therefore in agreement with this hypothesis. The fact that they are only found in a very small fraction of stars within a narrow spectral range centered near the spectral type of the sun is also in agreement with the ETI hypothesis. However, at this stage, this hypothesis needs to be confirmed with further work. Although unlikely, there is also a possibility that the signals are due to highly peculiar chemical compositions in a small fraction of galactic halo stars.

## 1. Introduction

A Fourier transform analysis of 2.5 million spectra in the Data Release 8 of the Sloan Digital Sky Survey (SDSS) and the SEGUE 2 SDSS was carried out to detect periodic modulations contained in their frequency spectra (Trottier 2012). The original motivation was to search for the periodic modulation caused by intensity pulses having short time separations (Borra 2010). These pulses could be generated by ultra-rapidly varying objects or other exotic sources. Borra (2010) carried out a theoretical analysis that shows that objects that emit intensity pulses separated by constant times shorter than $10^{-10}$ seconds generate periodic spectral modulations that are detectable in astronomical spectra. Note that the modulation is periodic in frequency units but not in wavelength units. The theoretical analysis of Borra (2010) is supported by experiments carried out by Chin et al. (1992). Searches for Extraterrestrial Intelligence (SETI) have been carried out since the first published suggestion by Cocconi & Morrison (1959). Tarter (2001) discusses the history of SETI including physical and sociological issues. There currently are searches being carried out with radio and optical telescopes. For example Korpela et al. (2011) discuss the UC-Berkeley SETI project. Borra (2012) suggested that extraterrestrial intelligence (ETI) could send pairs of light pulses separated by a constant time interval to generate periodic spectral modulations. ETI would use these signals to make us aware of their existence and send us messages.

The present article discusses some of the results of the Fourier transform analysis of SDSS spectra that found the type of periodic spectral modulation predicted by Borra (2010) in a small number of stars that are all within a narrow spectral range centered near the spectral type of the sun. We shall consider five possible physical causes of the spectral modulation: instrumental and data reduction effects, rotational transitions in molecules, the Fourier transform of spectral lines, rapid pulsations and finally the ETI signal predicted by Borra (2012),

## 2. Data analysis

The data analysis techniques are described in Trottier (2012) and in Borra (2013) and this section only gives a short summary. The expected periodic spectral modulations are periodic in frequency units (Borra 2010), consequently the SDSS spectra, which are given in wavelength units, must first be converted to frequency units. For the needed Fourier transform analysis, the frequency spectra must be sampled at equal frequency intervals, but the SDSS spectra are sampled at equal wavelength intervals and therefore are not located at equal frequency intervals. Consequently, after conversion to intensity units, the intensities are not sampled at equal frequency intervals. To obtain intensity samples at equal frequency intervals, we take the two intensity values, at the frequency location obtained from the original SDSS wavelength locations on both sides of the required frequency intervals, and compute the intensity at the required frequency by carrying out a standard linear interpolation between the two original SDSS intensity values. The frequency spectra are then analyzed with Fast Fourier Transform (FFT) software. One must use simple signal finding algorithms because the number of spectra to analyze is very large (2.5 millions). The FFT is performed on the difference between the frequency spectrum and a continuum spectrum, obtained from the smoothed spectrum, to remove a very strong and bumpy contribution at low times that would make it extremely difficult to detect a signal with automated software. The SDSS spectra typically contain 3900 digital samplings in the frequency domain and yield, after the FFT, 1950 independent samplings in the time domain. To make the sampling effect clearly visible, it is best to plot the Fourier modulus as a function of the sampling number $N$ instead of time units. Time increases linearly with $N$ and can be converted to seconds by multiplying $N$ by 2.1538 $10^{-15}$. This number can vary a little because the actual number can vary from spectrum to spectrum since it depends on the total number of samplings in the digitized SDSS spectrum, which vary, by a few samplings, from spectrum to spectrum. However 2.1538 $10^{-15}$ positions most of the detected objects within the time interval in Figure 2 and therefore gives a reasonable estimate.

To detect a signal, the software flags objects that have a peak in the FFT spectrum having a signal to noise ratio (*SNR*) greater than a preset value. A statistical analysis must use Rayleigh statistics because we use the Fourier modulus of the FFT. Rayleigh statistics have a cumulative distribution function given by

$$F(x) = 1 - e^{-x^2/2\sigma^2} , \qquad (1)$$

where $\sigma$ is the standard deviation. To compute $\sigma$ the Fourier spectrum is divided in 8 separate contiguous boxes of 250 $I_i$ samples. The standard deviation $\sigma$ is given by the Rayleigh statistics formula

$$\sigma^2 = \sum_{i=1}^{250} I_i^2 / 500 \quad , \qquad (2)$$

where $I_i$ is an intensity sample in the Fourier modulus spectrum. This value of $\sigma$ is used to detect a signal from the signal to noise ratio $SNR = I_i/\sigma$ at all locations within the box where it is evaluated. The contribution of the underlying "continuum", coming from the Fourier transform of the frequency spectrum minus the smoothed continuum, is included in the values of $I_i$ in Equation 2 as well as the values of the signal. Borra (2013) discusses the reasons why no attempts are made to remove the underlying Fourier "continuum". The underlying Fourier "continuum" contributes to the noise evaluated from Equation 2, and therefore decreases the *SNR*. On the other hand it has the opposite effect of increasing the signal. In practice, these problems are minor for the results discussed in section 3 because the contribution of the Fourier "continuum" is relatively small for $N > 500$, even for bright objects. At the end of section 3, which discusses the data analysis results, we show that they validate the use of Rayleigh statistics.

To avoid too large a number of detections, due to the very large number of objects analyzed multiplied by the 1950 independent samplings, that would have to be individually inspected, only peaks giving a $SNR > 5.5$ had to be used for $N > 250$ because there would otherwise have been too many detections of false signals coming from bumps in the Fourier spectrum (mostly for N< 500 and bright objects). These flaws of the detection method are not important at this stage of the search, where the main purpose is to find peculiar objects and a quantitative estimate of their occurrence rates is less important. The main purpose of the software is to flag interesting spectra which can then be further analyzed. Note however that, in practice, because of these flaws, the actual *SNRs* are greater than those measured by the software. This can be seen in the next section where we carry out a better evaluation of the *SNR* of the star in Figure 3.

## 3. Results

The analysis for $N < 250$ is difficult because the Fourier spectrum is very bumpy (see previous section) and is discussed in Borra (2013). For $N > 250$ the number of detections is consistent with the number expected from Rayleigh statistics but with a standard deviation that is slightly larger (8%) than the standard deviation computed for Rayleigh statistics (Equation 2). This is consistent with an overestimate of the standard deviation that comes from the contribution of the Fourier continuum as discussed in section 2. We could therefore have assumed that all the detections are due to the effect of noise; however if this were the case the detections would be uniformly distributed, within reasonable statistical fluctuation boundaries, as a function of the sampling number $N$. Figure 1 shows that this is not the case and that the excess of detections is due to a strong excess within the narrow range $750 < N < 800$. We can therefore conclude that a signal having a period within that interval is present in a small fraction of stars. A detailed analysis gave interesting results which are discussed below. We find that the signal is only present in a very small fraction of stars within a narrow spectral range centered on G0.

Figure 1 shows the distribution of detected signals as a function of the sampling number $N$ at which they are detected in the Fourier spectrum for stars only. Like in all figures that follow, the error bars show +- a standard deviation. It is immediately apparent that there is a considerable excess of signals in the box at $750 < N < 800$. Similar plots for quasars and galaxies did not show an excess of signals at such $N$ locations. The overwhelming majority of the detected signals (45 out of a total of 51) in the $750 < N < 800$ box are located at $N = 764$ or $N = 765$. The total number of samples in a SDSS spectrum varies by a few values among the spectra. Note also that different SDSS releases give spectra with slightly different numbers of samples so that the FFT may give peaks at slightly different $N$ locations than $N = 764$ or $N = 765$. When correcting for this variation, we find that the location of the peaks is even narrower since the peaks are actually located at the same $N$ position. This can be seen in Figure 2 which shows the location of the peaks after conversion to time units. The conversion to time units corrects for the total number of samples in a spectrum. In time units $\Delta N = 1$ corresponds to $\Delta t = 2.1538 \; 10^{-15}$ seconds, which is the total width of the horizontal axis in Figure 2. Consequently, the overwhelming majority of the detected signals in the box at $750 < N < 800$ in Figure 1 have actually the same $N$ location. Table 1 lists the stars that were detected with times between $1.6453 \; 10^{-12}$ and $1.6474 \; 10^{-12}$ seconds. It gives the plate identification and fiber numbers, the right ascensions and declinations (JD2000), the spectral type, g magnitude and the signal to noise ratio (SNR) of the detected signal. The plate ID and fiber numbers in Table 1 can be used to find additional information in the SDSS web site (http://data.sdss3.org/advancedSearch). In particular, the finding charts and u, g, i, r, z magnitudes are given.

Figure 3 shows the Fourier modulus of the frequency spectrum (after subtraction of its smoothed spectrum) of a star that had a statistically significant signal. The figure shows a signal at $N = 765$ that significantly stands out above the noise. Figure 4 shows a zoom in the region of the signal. The signal can clearly be seen in both figures. Figure 4 clearly shows that the signal has a width of 1 sample ($\Delta N = 1$). The Fourier spectrum in Figures 3 and 4 has the typical appearance of the Fourier spectrum of a detected star. The two plots in Figure 5 show the frequency spectrum of that star. The bottom plot shows the unaltered frequency spectrum. The sinusoidal signal is totally undetectable by eye inspection because it has very small amplitude. To make the characteristics of the signal visible we added to the frequency spectrum a sinusoidal signal that has the period of the detected signal but with amplitude multiplied by $10^4$. This spectrum is plotted at the top of Figure 5. The added signal looks like added noise because the period is very short. The spectrum in Figure 3 was flagged, by the software that analyzes all spectra, with a signal to noise ratio $SNR = 6.09$. However, as we write in section 2, the evaluation of the $SNR$ by the software has flaws, that come from the difficulty of subtracting the Fourier continuum with automated software in an extremely large number of objects, and give an $SNR$ that is smaller than the actual $SNR$. We evaluate again the $SNR$ of this object to show the effect of these flaws. Because the periodic signal in Figures 3 and 4 is superposed to a background value equal to the continuum of the underlying Fourier spectrum, we can subtract this background and we then find that the net signal has an amplitude of $I_{765} \sim 90$. Because the Fourier transform of white noise (photon noise in our case) also gives white noise, we can evaluate the standard deviation at large values of $N$, where the contribution from the Fourier continuum is totally negligible (as seen in Figure 3 for $N > 1000$), and then apply it for an evaluation of the signal at $N = 765$. The white noise standard deviation is $\sigma = 8.0$ for that spectrum, giving a $SNR \sim 11$ and, using Equation 1, a probability of detection, due to the noise only, of $5 \cdot 10^{-27}$. Considering that over 2 million SDSS spectra were analyzed this gives a probability $\sim 10^{-20}$ that the signal in the Fourier spectrum of an object at a given $N$ location is due to statistical fluctuations. The real $SNR$ is therefore clearly significantly larger than the *SNR* at which it was flagged (*SNR = 6.09*). This is the case for all of the stars in Table 1.

Figure 3 shows the Fourier transform of an F5 star with a statistically significant signal in the $762 < N < 765$ range. Figure 6 shows the Fourier transform of an F star template taken from the SDSS database. It does not show any significant signal near that range. The Fourier transforms of templates of other spectral types also did not show the signal. Figure 7 shows the Fourier transform of an F5 star that did not have a statistically significant signal. Like the FFT of the template, it does not show any significant signal near that range. To show examples of stars with different spectral types that have the signals, Figures 8. 9, 10 and 11 respectively give Fourier transforms of a K1, G2, another F5 star, and an A0 star that had the signal. They clearly show the same type of signal, at the same location, seen in Figure 3. Figures 3 to 11 were selected at random among the stars that had a signal or no signal simply to visually illustrate the nature of the detections or lack of detections.

Note that the discussion that follows in the present and the next paragraphs of section 3 is very important for it clearly shows that the signal cannot be due to data analysis or instrumental effects. Figure 12 shows a histogram of the distribution of the number of stars that have a periodic spectral modulation signal as a function of spectral type. We see that the majority of the signals are present in spectral types ranging from F5 to G2 and that neither galaxies nor quasars were detected. However, one must worry about selection effects in the distribution of the stellar spectra observed in the SDSS data used. The SDSS data release 8 used to generate Figure 12 includes SEGUE (Sloan Extension for Galactic Understanding and Exploration) spectroscopic data. The SEGUE survey targeted stars that are mostly F to K stars so that there clearly is a bias. For comparison, Figure 13 shows a histogram of the distribution of all the spectral types in the SDSS data that we analyzed, including galaxies. For a proper comparison with Figure 12 only spectra of stars and galaxies having a median spectrum *SNR > 40* are included in Figure 13. This limit was chosen because 90% of the detected stars have a median spectrum *SNR > 40*. The median spectrum *SNR* used is the number given with the SDSS spectra in the SDSS web site to quantify the median *SNR* of the spectrum. This limit is particularly necessary for galaxies which are, on average, fainter than stars. The number of quasars above this limit is negligible and they are not included in the figure. The spectral types are the spectral types identified in the SDSS, where they are quantized in the intervals plotted in Figures 12 and 13. This explains why they are quantized in particular spectral type. For example all G stars in the SDSS survey are only classified as G0, G2 or G5 and all A stars are classified as A0. Figure 13 shows that there is a selection effect in the spectra that were observed. However, a comparison of Figure 12 and 13 shows that a significant number of galaxies, and stars later than K1 are present in Figure 13, while none were detected in Figure 12. Also, the numbers of A0 stars detected are significantly below the number expected by comparing Figures 12 and 13. Because galaxies are extended objects, one might suspect that the lack of detections is caused by their extendedness. However this cannot be, because the SDSS spectra are obtained with optical fibers that have a 3 arcseconds diameter, which is comparable to the average diameter of stellar images caused by the atmospheric seeing. The comparison of these 2 figures therefore confirms that, notwithstanding the selection effects in the survey, the peaks are mostly present in the F2 to K1 spectral range.

We then carried out Fourier transforms of the spectra in the SEGUE 2 survey, which contains the spectra of 120,000 stars. The *SNR* threshold of detection was lowered to 5.0. We did not carry out Fourier transforms with a lower threshold of detection because there would have been too much detection of objects without a signal caused by the noise. Like in Figure 1 and 2, the majority of the peaks are detected in the same small period range. Table 2 lists the stars that were detected with times between $1.6453 \ 10^{-12}$ and $1.6474 \ 10^{-12}$ seconds (like the stars in Table 1). Figure 14 shows the histogram of the distribution of the detected peaks as a function of spectral type. Like in Figure 12 the majority of the signals are detected in spectral types ranging from F5 to K1. Figure 15 shows a histogram of all the spectral types in the SEGUE 2 survey data that we analyzed. Like we did for a comparison of Figures 12 and 13, we only include spectra that have a median spectrum $SNR > 40$. We see that there are a significant number of A0, K3, K5 and K7 stars in Figure 15 while very few A0 and K3 were detected and no K5 and K7 were detected in Figure 14. Note in particular that in Figure 15 the sum of the number of K5 and K7 stars is 73 % of the number of F9 stars. Consequently we would have expected to have detected about 57 stars while none were detected. Furthermore, the number of A0 stars in Figure 15 is 24 % the number of F9 stars, so that we would have expected to detect about 19 A0 stars, while only 1 was detected. Finally, the number of K3 stars detected is also below the numbers expected. Figure 15 shows that the number of K3 stars is half the number of F9 stars so that about 40 stars should have been detected while only 6 were detected. These lacks of detections have a very high statistical significance, as discussed in section 4. This lack of detections can also be seen by comparing Figure 15 to Figure 16 which shows the number of detected stars normalized to the number of detected F9 stars, using the relative distribution of spectral types in Figure 15. Figure 16 shows that the detections are consistent with a box-like distribution between F2 and K1. It is remarkable that the box is centered on G0, therefore very close to the spectral type of the sun.

As discussed in section 2, our statistical analyses that use signal to noise ratios use Rayleigh statistics with a cumulative distribution function given by Equation 1 because we use the Fourier modulus of the FFT. The continuous line in Figure 17 gives the theoretical Rayleigh distribution that we use, while the filled circles give the fraction of the number of detections of peaks in stars. For a proper comparison, we do not include the stars with peaks detected in the range $762 < N < 765$, because these are peaks in the narrow range where we have found real signals. We can see a good agreement between the theoretical and observed numbers if we consider the uncertainties coming from the small numbers of detections.

In conclusion, the analysis of the SDSS data release 8 and SEGUE 2 spectra show that the signal is only present in a very small fraction of stars within a narrow spectral range centered on G0.

We shall consider possible causes of the periodic spectral modulation. The periodic modulation could come from instrumental or data analysis effects, the spectra of rotational transitions in molecules or the Fourier transform of spectral lines that have nearly periodical positions in the frequency spectrum. It could come from an exotic physical phenomenon. Finally, it could also come from signals sent by extraterrestrial intelligence, which was one of the two main reasons for Fourier analyzing the SDSS spectra.

## 4. Instrumental and data analysis effects

Instrumental or data analysis effects could generate the detected signals. The detailed quantitative analysis that follows shows that this hypothesis is not valid.

To begin, let us note that the Fourier signals are very weak. Typically, they contain of the order of a few times $10^{-5}$ of the total energy of the spectrum. Prima facie this number seems small considering the fact that Figure 4 and Figures 7 to 11 show that the Fourier signals are significantly above the noise in the Fourier spectrum. However one must first note that the Fourier transform can extract a weak periodic signal buried within the noise in the frequency spectrum because it contains all of the energy contained in a signal that is spread over a large frequency range in the frequency spectrum. Secondly, our FFTs are carried over the difference between the frequency spectrum and the smoothed continuum spectrum (see section 2) so that the very large contribution of the continuum is not included in these figures.

The SDSS spectra are obtained by a pair of spectrographs fed by optical fibers that simultaneously measure more than 600 spectra with a single observation in the three-degree field of the telescope. The Fourier signals would have an instrumental origin if the objects that have the signals were always located in the same fibers. The fibers are identified by numbers in the individual spectra of the SDSS data and we find that the detected objects are located at random among the fibers. There is also no dependence on the position of the field of view. This can be seen in Tables 1 and 2 that list the detected objects and give the plate and fiber numbers. Another possible origin of the signals may come from instrumental problems that may only be present during a short time. This could, for example due to instrumentation maintenance and upgrade. In that case the detections would be present in a small number of plates obtained within small time frames. However Table 1 gives the Julian dates (MJD) and shows that the signals were detected over a time interval of 3020 days (8.3 years) from MJD 2451612 to MJD 2454653. This is essentially the entire duration of the Data Release 8 of the SDSS spectroscopic survey that we used. Only 4 of the plates (out of 42) have detections in 2 fibers of the 600 fibers that are positioned on objects in a single plate. The remaining 38 plates all only have a single detection per plate. It is important to be aware that the Data Release 8 survey used to generate Table 1 was mostly dedicated to obtain spectra of quasars and galaxies (of which some candidates turned out to be stars) but also includes observations from the SDSS SEGUE 1 survey which obtained spectra of 240,000 stars to create a detailed three-dimensional map of the Milky Way. The SEGUE 1 and SEGUE 2 surveys have spectroscopic selection effects which are discussed in section 3. The SEGUE 1 data (in Table 1) were obtained within a time span going from the winter 2004 (MJD 2453000) to the winter 2008 (MJD 2454600 ). These are the approximate values of the time span given in the SDSS web site. The fact that most of the stellar spectra in the Data Release 8 survey are stars in the SEGUE 1 survey explains why the majority of the MJDs of the detections in Table 1 are within these MJD intervals. Table 2 contains results from the SEGUE 2 that spectroscopically observed around 120,000 stars, focusing on the stellar halo of the Galaxy, from distances of 10 to 60 kpc. These observations were obtained in a small time interval from the Fall of 2008 (MJD 2454730) and the Spring 2009 (MJD 2455000). Table 2 shows that the Julian dates of the detected spectra are spread within that MJD interval. The web site of the SDSS surveys discusses in greater details how the data were obtained. The SEGUE 1 and SEGUE 2 surveys obtained stellar spectra within the spectral range A0 to K7 as can be seen in Figure 13 for the SDSS data release 8 and figure 15 for the SEGUE 1 survey. The impact of these selection effects on the detections are discussed in details a few paragraphs below.

A major characteristic of the detected signals come from the sharpness of the Fourier signals, which always are one sampling wide ($\Delta N = 1$), as can be seen in Figure 4. This sharpness signifies that the periodic modulation covers the entire spectral range. To further validate this fact, we carried out separate Fourier transforms for the blue (380 nm $< \lambda <$ 600 nm) and red regions (600 nm $< \lambda <$ 9200 nm) of spectra of the objects in Table 1. The spectra in these two regions were obtained by two different spectrographs. The Fourier transforms were carried out with the Non-Uniform Discreet Fourier Transform software discussed in the paragraph that follows the next paragraph. Figure 18 shows the Fourier transform of the blue region of the same spectrum that was used to generate Figures 3 and 4 while Figure 19 shows the Fourier transform of its red region. Because we use, in both regions, half of the spectral range used to generate Figures 3 and 4, the intensity of the Fourier signal is decreased by a factor of 2. As discussed in section 2, the noisy background that one sees in the Fourier modulus at the location of the signal comes partly from photon noise and mostly from the Fourier transform of the frequency spectrum minus the smoothed continuum. The noise coming from the frequency spectrum is not decreased as strongly by dividing the region in two. This type of noise has a bumpy structure for low values of $N$, as can be seen in Figure 3. Because the signal in Figures 18 and 19 is closer to $N = 0$ than it is in Figure 3, since we only use half the spectrum, the bumpiness is greater than in Figure 3 and this is why there is a strong very noisy background in Figures 18 and 19. Consequently, the signal is very difficult to see if we plot the entire Fourier modulus of the blue and red regions and this is why Figures 18 and 19 only display a small range in the region of the signal. We do not use the sampling numbers because they do not have the same sampling numbers as in Figures 3 and 4 since they use half the spectrum and therefore have half the samplings of the entire spectrum that generates Figures 3 and 4. We plot instead the modulus as function of the period in seconds to show that both the blue and red regions give signals within the spectral range in Figure 2, which gives the locations of the detected peaks converted to time units. Figures 18 and 19 show peaks well within the range of the detected signals in Figure 2. Although the signal stands out less in Figure 19 than it does in Figure 18, this is simply because the background noise is stronger in Figure 19. The signals have similar amplitudes above the background in both figures. This type of periodic spectral modulation could come from the interference between two beams originating from a single beam which is split in two components that are then recombined after a path difference between the beams is introduced by the instrumentation. Because interference is constructive if the optical path difference is equal to an integer multiple of the wavelength, and destructive at an integer plus 0.5, there is a wavelength dependent modulation of the intensity of light. This effect is discussed in optics textbooks, usually in the context of thin-film interference that causes colorful interference patterns in thin films. This is the type of interference that comes from the partial reflections of a beam at two glass interfaces separated by a distance *d*. The time separation between the 2 beams, given by the Fourier transform, is then $t = d/(cos(\alpha)c)$, where $\alpha$ is the angle of the incidence of the beam on the interfaces and *c* is the speed of light. Assuming that the signals are due to the interference between two beams generated by reflections at two glass interfaces, we find that the signal at *N = 765* seen in Figure 4 could be produced by two partial reflections at two interfaces separated by a distance *d <* 250 microns, the upper limit corresponding to beams perpendicular to the interface. Prima facie, the *d <* 250 microns separation value seems like a very small value for the separation between 2 reflecting surfaces. Furthermore, as discussed below in this subsection, the signal contains only a few times $10^{-5}$ the energy of the spectrum. This seems like too small a value for this type of interference since it implies two surfaces having reflectivity values of the order of $10^{-5}$.

The SDSS spectra are obtained from two different spectrographs, one in the blue region (380 nm < $\lambda$ < 615 nm) and one in the red region ( 580 nm < $\lambda$ < 920 nm ), while the signals extend over the entire spectral range. Intuitively one would not expect two spectrographs, detecting in two different spectral regions, to give a signal having exactly the same period. The hypothetical interfering beams would therefore more likely originate in the optical elements that send the light to the spectrographs. These optical elements are rather simple, consisting of a collimator mirror and a beam splitter. However, this interference effect should be present in all objects, irrespective of spectral type, contrary to what we find since only a very small number of objects within a narrow spectral range have the signal. Furthermore, as discussed in the previous paragraph, the small separation (<250 microns) and reflectivity ($10^{-5}$) needed to generate the signals appear unrealistically small.

As discussed in section 2, we converted the SDSS spectra from the original equally spaced wavelength locations to equally spaced frequency locations, using a linear interpolation between the original SDSS locations, and one may wonder whether this conversion would cause the detected periodic signals. This would be a data analysis effect and, as discussed in details in the present section and summarized in the abstract: The signals cannot be caused by instrumental or data analysis effects because they are present in only a very small fraction of stars within a narrow spectral range and because signal to noise ratio considerations predict that the signal should mostly be detected in the brightest objects, while this is not the case. To further verify the effect of the conversion from wavelength to frequency, we computer simulated SDSS spectra. The spectra were generated in wavelength units by creating a continuum in the wavelength interval of the SDSS spectra (380 nm < $\lambda$ < 920 nm) to which we added 105 spectral lines modelled by Gaussian functions having a width that gives the average resolution of the SDSS survey ($\lambda/\Delta\lambda$= 2000). The lines were located at random locations and had intensities that varied at random. The spectra were sampled at the wavelength locations of the samplings of the SDSS spectra. These spectra were then analyzed with the same software used to analyze the SDSS spectra. The Fourier transforms did not find any signals and gave plots of the Fourier modulus similar to those in Figures 6 and 7. The fact that the conversion from wavelength to frequency does not cause a false signal is also confirmed by Figure 6, which shows the Fourier modulus of an F star template taken from the SDSS database. There is no signal in Figure 6. The Fourier transforms of templates of other spectral types also did not show the signal. To further verify the effect of the linear interpolation, we Fourier analyzed the spectra in Table 1 with software that does not use the linear interpolations. We carried out the Fourier transforms on spectra as a function of frequencies located at the SDSS wavelength locations converted to frequencies. We used publically available versions of software that use a variety of numerical techniques of Non-Uniform Discreet Fourier Transform (NDFT) to carry out Fourier transforms in data that are not uniformly sampled. They gave Fourier transforms that were essentially the same as those obtained with the FFT and linear interpolations. They found the signal at the same location with only negligible differences in intensity. Figure 20 shows the Fourier transform obtained with a NDFT of the spectrum used to generate Figure 3. A comparison of the two figures clearly shows that the differences are negligible. Figure 20 has a peak at the same location as the peak in Figure 3, with a strength that is only slightly different.

The fact that no galaxies were detected and that the detected stars are within a narrow spectral type range centered around G0 is one of the two major facts that go against the hypothesis of an instrumental or data analysis effect since it should be present in all stellar spectral types and in Galaxies. Note that the peaks detected in the Galaxies in Borra (2013) were detected at totally different locations ($40 < N < 65$) that increased linearly with redshift along a very tight relation (see figure 3 in Borra 2013) and have therefore no relation with the $N \sim 765$ peaks detected in stars. Figure 16 shows that the signal is not present in stars later than K3 and that it is present in only a very small fraction of A0 and K3 stars, while instrumental or data analysis effects should be present in all spectral types. A comparison between figures 12 and 13, which use SDSS data that do not include the SEGUE 2 SDSS data in Figure 16, also shows that the signal is only present in stars having the same spectral range and that no galaxy has the signal. This selection effect has a very high statistical significance, as can clearly be seen by considering the results obtained from the SEGUE 2 survey that we discuss in the next paragraph.

Figure 15 shows the distribution of all the spectral types in the SEGUE 2 survey, where, for a proper comparison, only stars having a median signal to noise ratio of the entire spectrum comparable to the stars having a detected signal are included. We can obtain an estimate of the number of stars of a spectral type that should have a signal by using the ratio between the number of stars of that spectral type in Figure 15 and the number of stars having an F9 spectral type in Figure 15 and then multiply this ratio by the number of F9 stars having a detected signal (Figure 14). The reason why we normalize with respect to F9 can be seen in Figures 14 and 15, which show that F9 is the spectral type that has the most detections. We then find that stars in the K5 to K7 spectral range should have 57 detections, while none were found. This number of detections of K5 and K7 stars is 7.5 standard deviations away from the expected number, with a probability of $3.2 \times 10^{-14}$ that it is due to random number fluctuation. There should have been 19 detections of A0 stars, while only 1 was detected. This detection of 1 A0 star is 4.2 standard deviations away from the expected number, with a probability of $1.3 \times 10^{-5}$ that it is due to random number fluctuation. The combined number of detections of A0, K5 and K7 stars is 8.5 standard deviations away from the expected number, with a probability less than $10^{-16}$ that it is due to random number fluctuation. Furthermore 39 K3 stars should have been detected while only 7 were detected. Clearly the selection effect cannot be due to random fluctuations. Note also that a comparison of Figures 12 and 13 that come from SDSS survey data that do not include the objects in the SEGUE 2 survey, confirms this conclusion. Using the same analysis but using the data shown in Figures 12 and 13, we see that we should have detected 7 galaxies (0 detected), 6 A0 stars (2 detected), and a total of 7 stars in the range K3 to M (0 detected). These numbers are, respectively: 2.6, 2.0, and 2.6 standard deviations from the expected numbers.

The second major fact that goes against the hypothesis of an instrumental or data analysis effects comes from the analysis that follows that shows that only about 1% of the stars having a spectral type between F2 to K1 have the signal, while an instrumental or data analysis effect should be present in all objects. The black dots, at the very bottom of Figure 21 give the fraction of detected stars, obtained by dividing the number of detected stars by the number of observed stars, as a function of the median signal to noise ratio of the spectrum. For a proper comparison, we only have taken into account the number of observed stars within a spectral range from F2 to K1 because the signals were overwhelmingly detected within that spectral range. We see that a signal is detected in a very small fraction of stars within that narrow spectral range. The two dashed lines show the region surrounding the black dots within +/- a standard deviation. Assuming that the signal is present in all stars within a spectral range from F2 to K1, as it would be the case for an instrumental or data reduction effect, this would be due to the fact that the signal is very weak so that the signal to noise ratio is small and therefore only a small fraction of the signals are detected. There then should be an increase of detections with the increase of the median signal to noise ratio of the spectrum if we make the assumption, valid in the case of instrumental or data reduction effects, that the signal to noise ratio of the signal should increase proportionally to the median signal to noise ratio of the spectrum. The continuous line in Figure 21 shows the increase of the fraction of detected stars with increasing median spectrum SNR of the spectrum predicted by Rayleigh statistics. To compute the continuous line we assume that the signal is present, and has the same amplitude, in all stars, like it would be the case if the signal came from instrumental or data reduction effects. We then assume that it is only detected in $1.4 \cdot 10^{-3}$ stars at a median signal to noise ratio of the spectrum of 34 because it is a weak signal that is detected only for statistical reasons because noise effects increase the signal 4.0 standard deviations above its true value. We then see that the continuous line in Figure 21 predicts a large increase of detections with increasing median signal to noise ratio, while the fraction of detected stars remains at a nearly constant very small value. We can therefore draw two important conclusions. On the one hand, this shows that the hypothesis that there is a very weak signal in all stars is wrong. On the other hand, this is what is expected if a stronger signal is present only in a very small fraction of stars. The next paragraph quantitatively evaluates these conclusions.

The black dots in Figure 22, like in Figure 21, give the fraction of detected stars, obtained by dividing the number of detected stars by the number of observed stars within a spectral range from F2 to K1, as a function of the median signal to noise ratio of the spectrum. The two dashed lines show the region surrounding the black dots within +/- a standard deviation. The large standard deviations come from the small number of detections. There only are a total of 180 detections and the significant increase of the standard deviations for a median *SNR > 70* comes from the fact the number of detected and observed stars decreases rapidly for a median *SNR > 70*. The fraction of detections increases with the median *SNR* up to *SNR* = 50 and remains constant in the range between *SNR* = 50 and *SNR* = 70. The increase of detections above *SNR* = 80 is not statistically significant since it is within the significantly larger uncertainties. The continuous line in Figure 22 shows the increase of the fraction of detected stars predicted by Rayleigh statistics if we assume that there is a signal, significantly stronger than the signal assumed in Figure 21, that is present in only 1% of the stars. Consequently, at a median signal to noise ratio of the spectrum of *SNR* = 34, the signal is actually detected in $1.4 \cdot 10^{-1}$ stars that have a signal, and the signal is detected only for statistical reasons because noise effects increase the signal 2.4 standard deviations above its true value. We see a reasonable fit within the statistical uncertainties given by the dashed lines. The fraction of detections is therefore consistent with a signal that is only present in approximately 1 % of the stars that have a spectrum within a spectral range from F2 to K1. This is not what is expected from an instrumental or data reduction effect, since the effect should be present in all of the stars and all of the spectral types. This has also implications in the context of signals from ETI discussed in section 5.4.

One may also wonder about the magnitude distribution of the objects since an instrumental effect may be magnitude dependent. Tables 1 and 2 give the magnitudes of the detected objects. They show g magnitudes mostly in the range $15.5 < g < 18$, with a few objects below and above this magnitude range. The objects detected are from the SEGUE survey that obtained spectra in the range $14 < g < 20$ (Yanny et al. 2009). Tables 1 and 2 therefore show that there is no strong magnitude effect. The smaller number of detections at bright and faint magnitudes should be expected since there are only few objects at bright magnitudes and the signal to noise ratio will be low at faint magnitudes and make a signal harder to detect. Note also that a few objects with $g < 15.5$ and $g > 18$ were detected. The median signal to noise ratio of the spectrum, obtained from the SDSS database is a much better measure of the brightness of the objects detected. The black dots, at the very bottom of figures 21 and 22, give the fraction of detected stars as a function of the median signal to noise ratio of the spectrum. They show that the fraction of detections is nearly constant. This is discussed in section 4, where a statistical analysis comes to the conclusion that this goes against the hypothesis of an instrumental or data analysis effect.

We can therefore conclude that the signals are not caused by instrumental or data analysis effects.

### 5. **Signals from the Fourier transform of spectral features**

The signals could come from the Fourier transforms of spectral features. However, this hypothesis must explain why the signals are detected in only a very small fraction of stars within a small spectral range. We shall consider two possibilities; that they are caused by the Fourier transforms of rotational transitions in molecules or the Fourier transforms of spectral lines.

*5.1 Rotational Transitions in Molecules*

Rotational transitions in molecules can give lines that are equally spaced in frequency and could therefore give a signal similar to the signal seen in Figures 3 and 4. However, molecular spectra are mostly in the infrared spectral region. In the spectral range of SDSS spectra (380 nm to 920 nm) the equally spaced transitions in molecules generate spectral lines that are contained within narrow spectral bands, while the small width of the detected peaks indicates that the periodic modulation should be present in most of the spectrum (see section 3). The molecules cannot be in the interstellar medium because they would be present in all spectral types and in galaxies while it is not the case (see section 3). Grevesse & Sauval (1994) and Sinha (1991) list the molecules observed in the spectrum of the Sun that have spectral lines within our spectral range. The SDSS spectra have a very low spectral resolution *($\lambda/\Delta\lambda \sim 2000$)* and low *SNR*, consequently the lines detected in the SDSS spectra should be strong lines present in the sun and therefore the list in Grevesse & Sauval (1994) and in Sinha (1991) should contain them all. An examination of the list of these spectral lines in databases of molecular spectra (http://www.cfa.harvard.edu/hitran/ , Rothman et al. (2009) and http://kurucz.harvard.edu) did not find spectral lines having the required frequency separations.

*5.2 Fourier Transform of Spectral Lines*

The signals could come from the Fourier transform of the intensity profiles of spectral lines that are nearly equally spaced in frequency. Borra (2013) carries out a discussion of this effect and we shall apply a similar analysis to our signals. The purpose of this modelling is to determine the general characteristics (e.g. number and amplitude) of the spectral lines needed to generate the Fourier signals that we have found and then compare them to the data to see whether these theoretical parameters are in agreement with the data. We shall start from a Shah function model convolved with a Gaussian because it gives periodically spaced lines, consequently this is a convenient model for the main feature of the Fourier signals that we found, which indicate a periodic signal in the frequency domain. Borra (2013) starts by modeling the frequency spectrum of absorption lines with a Shah function (also called a Dirac comb) *III(f-nF)* , where *f* is the frequency, *n* is an integer number, and *F* is the period of variation in frequency units, convolved with a Gaussian having dispersion $\sigma$ that models the intensity distribution with frequency *f* of a line. The Fourier transform of the convolution of two functions is equal to the product of the Fourier transforms of the two functions; consequently the Fourier transform of the Shah function *III(f-nF)* convolved with a Gaussian having dispersion $\sigma$ gives the Shah function *III(t-n/F)*, having period *1/F* in the time domain *t* , multiplied by the Fourier transform of the Gaussian line profile, which is a Gaussian having dispersion $\sim 1/\sigma$. The Fourier transform of this model of the frequency spectrum gives a Shah function multiplied by a Gaussian centered at *t* = 0 (corresponding to *N* =0 in Figure 3), which is a Shah function with combs having an amplitude that decreases with increasing *N*. Figure 9 in Borra (2013) shows an example of this type of Fourier spectrum.

The lines of the F5 stars, like the star used to generate Figures 3 and 4, have, in SDSS spectra, full widths at half maximum of 5 $10^{11}$ Hz (equal to 4 samples in a SDSS spectrum), Note that these full widths are the widths in the SDSS spectra and therefore are dominated by the instrumental resolution. They can be modeled by a Gaussian having dispersion σ = 2.12 $10^{11}$ Hz. The Fourier transform of this Gaussian in frequency units gives a Gaussian in time units having a dispersion of 8 $10^{-13}$ seconds which corresponds *to N* = 400 in Figure 3. The position of the detected signal (*N* = 765) is therefore at a distance of nearly twice the dispersion, which would imply that it would have a strength 4 times smaller than a signal located at *N* = 400. Looking at Figure 3, we see that there is no signal near *N* = 400 that is 4 times stronger than the peak at *N* = 765. This would signify that the signal at *N* = 765 is the first tooth of the Shah function and that the frequency spectrum has 729 lines separated by a constant frequency spacing of 6.127 $10^{11}$ Hz. This theoretical model has obvious flaws because it has absorption lines having equal intensities equally spaced in frequency. However it is useful to have an intuitive understanding of the situation and to understand the analysis that follows which starts from this model.

Using the methods described in Borra (2013), we improved the Shah-function model with computer simulations that change at random the intensities of the lines as well as the frequency locations of the individual lines predicted by the Shah function. The effect on the Fourier transform of this improved model is to decrease the strength of the peaks at higher *N* with respect to the strength of the first peak. This strengthens the conclusion in the previous paragraph that the detected signal must be the first tooth of the Shah function. This model therefore allows us to find the period *F* in the frequency Shah function *III(f-nF)*, which gives the separation between lines and therefore the total number of lines needed. Figure 23 shows the frequency spectrum obtained from such a simulation which starts from a Shah model with 729 lines separated by $6.127 \; 10^{11}$ Hz. The intensity of the lines and their central frequencies are varied at random with the technique described in Borra (2013). The average deviation from the periodic frequency position is +- 0.11 %. Figure 24 shows the FFT of the spectrum in Figure 23. The noisy Fourier spectrum at values of *N* < 600 in Figure 24 is caused by the Fourier transform of the absorption lines. Figure 24 only shows one peak because the other peaks, all at *N* > 800, are considerably weakened by the deviations from the Shah periodicity, as explained in the previous paragraph, and are below detection. The simulated lines in Figure 23 would be weak lines in the SDDS spectra with amplitudes comparable to the noise and therefore undetectable. However, they would be detected as strong lines in better quality spectra that have a higher spectral resolution because they have an average strength of the order of 1/20 of the strength of the Lyman $\alpha$ line in F5 stars. The simulations therefore show that the detected stars should have an extremely large number of strong spectral lines located at very nearly periodic frequency locations that should be clearly detectable in high-quality spectra; however the same lines should be absent or considerably weaker in most stars. Consequently, on the basis of this model, we can conclude that an unrealistically large number of lines positioned at nearly periodic frequency positions are needed. This large number of peculiar lines would only be present in 1% of the stars, implying that there is 1% of highly peculiar stars in the F2 to K1 spectral range. Note also that the number of lines needed (729) is comparable to the total number of samples in a SDSS spectrum (3900) that has a spectral resolution $\lambda/\Delta\lambda$ = 2000, which makes the hypothesis even less credible. Furthermore, one would expect the strength and the number of the lines to vary with spectral type, which should show a correlation of peak strength with spectral type, while, as discussed in section 6, the Fourier transform spectra of the detected stars did not show such a correlation.

      We can therefore conclude that the hypothesis that the signal is caused by the Fourier transform of spectral lines is unlikely. In the next section we reconsider the hypothesis that the signals come from the Fourier transform of spectral lines and consider possible effects that

might change the shapes and locations of the spectra lines and therefore might be responsible for the small fraction of detections given by the black dots in figures 21 and 22.

6. **Consideration of effects that change the shapes and locations of the spectral lines**

We will start from the hypothesis that the signals come from the Fourier transform of spectral lines.

The most peculiar characteristic of the signals comes from the fact that they were found in a very small fraction of stars in the F2 to K1 spectral range. This could be caused by effects that change the shapes and locations of the spectra lines and therefore change the strength of a signal coming from the Fourier transform of the lines. These effects could come from spectral variations due to differences in chemical composition or differences in effective temperature or gravity. They could also come from the effects of different rotational velocities, turbulence or radial velocities. We shall begin by evaluating the strength of the hypothetical effect that is needed to generate the peculiar characteristics of the signals mentioned at the beginning of this paragraph. As we shall see, a very strong effect is needed.

We start from the assumption that the line strengths are the same among all of the stars within a spectral range and that some effect changes the shapes and locations of the spectra lines and therefore weakens the signal coming from the Fourier transform of the lines. The black dots at the bottom of Figures 21 and 22 show the increase of the fraction of the stars with a detected a signal as a function of signal to noise ratio (*SNR*). The effect weakening hypothesis must reproduce the small increase of the detections with *SNR* given by the black dots. The problem with this hypothesis is that it is difficult to reproduce, with it, the small increase of detections with *SNR* given by the black dots in Figures 21 and 22. To understand this, consider the continuous line in Figure 21. It gives the increase of the number of detections with signal to noise ratio predicted by Rayleigh statistics (which must be used for the modulus of the Fourier transform) if we assume that all the stars have the signal. We can see that Rayleigh statistics predict a very strong increase with *SNR*, which is in total disagreement with the observations given by the black dots. On the other hand, the continuous line in Figure 21 assumes that all the stars have a signal having the same strength. If we now consider stars that have weaker signals, they would generate other lines having the strong increases with *SNR* predicted by Rayleigh statistics, but starting at different values of *SNR*. We then would have to add the contributions of all the lines to generate the predicted increase with *SNR*. This would still generate a large increase with *SNR* similar (albeit not exactly the same) to the increase given by the continuous line in Figure 21.

While the qualitative discussion in the previous paragraph gives an easy to understand analysis, it is too simple because the average slope of the continuous line in Figure 21 decreases with a signal that becomes progressively stronger. We can easily understand this by considering an extremely strong signal that is not significantly weakened by the noise in the spectra. Obviously it would be detected in all the spectra and generate a horizontal line in Figures 21 and 22. Consequently, to agree with the black dots coming from our observations, which show a very small increase with S/N, one would need a relatively strong signal in all stars that is only slightly weakened in 1% of the stars but strongly weakened in the remaining 99%. A hypothetical signal decreasing effect would then have to introduce a very strong weakening for most of the other 99% of the stars since their contributions should not increase significantly above the black dots. We shall now carry out a quantitative evaluation of this.

Let us assume that the signal is present in all the spectra and that a hypothetical signal weakening effect is responsible for the observed small increase of detections with *SNR* given by the black dots. The continuous line in Figure 22 was computed assuming that only 1% of the stars within a spectral range from F2 to K1 have a strong signal, while none of the other 99% have it. We see a good fit within the +/- one standard deviation region surrounding the dots. Consequently, to agree with the data, one would need a signal weakening effect that causes a rapid decrease of the signal with increasing strength of the effect. We can do a modelling of this decrease by assuming that 1% of the stars have the signal strength used to generate the continuous line in Figure 22 and then add to the continuous line the contribution of the signals coming from the stars that have a weaker signal. We can simply model this contribution by assuming that the distribution of the number of stars as a function of the signal strength follow a box-like distribution made of 20 boxes that each contains 5% of the remaining 99% of the stars. To begin with, let us consider the contribution coming from the first box, which only contains the 5 % of stars that have a signal that is the least weakened. The dashed lines in Figure 25 show the results of this modelling. The dashed lines were generated by assuming that 1% of the stars have the signal that generated the continuous line in Figure 25 and then adding to it the contribution of 5 % of the stars that have a weaker signal. The continuous line in Figure 25 is the same as in Figure 22. The top dashed line shows the results in the case where the contribution comes from a signal that is weakened by a factor of 1.5, the middle dashed line shows the results for a signal weakened by a factor of 2 and the bottom dashed line shows the results for a signal weakened by a factor of 3. We can see that decreasing the contributing signal by a factor of 1.5 as well as a factor of 2.5 predicts a strong increase of detections with *SNR* that is in total disagreement with the observations. On the other hand, a decrease by factor of 3 only predicts a small increase that is within the statistical errors. The contributions from the stars in the next 24 boxes, which would contain weaker signals, would therefore have to be totally negligible. This simple model therefore shows that the hypothetical weakening effect responsible for the small number of detections would have to cause a signal weakening that increases very rapidly with the increase of the effect, since the contribution of the remaining 95% of the stars would have to be totally negligible..

    The conclusion of the modelling carried out in the previous paragraphs (that the hypothetical effect would have to cause a signal weakening that increases rapidly with the increase of the effect) is a very important fact to keep in mind while reading the next paragraphs that discuss the effects of differences in chemical composition, effective temperature, gravity, rotational velocities, turbulence and radial velocities.

    Prima facie, differences in chemical composition seems like a logical explanation. To begin let us note that, as discussed in section 5.2, a very large number of lines at nearly periodic locations are needed to generate the signals. The discussion in the previous paragraphs leading to Figure 25 also shows that extremely large differences in chemical composition are needed to explain the small increase of detections shown by the black dots in Figure 25, since the 1% of the stars with a signal would need much stronger lines than the majority of the other stars. Major differences in chemical compositions are not known to exist among stars in the F2 to K1 spectral range where the signals were detected. It is highly unlikely that these chemical compositions differences have never been found because 1% of the stars in F2 to K1 spectral range have the signal and therefore 1% of the stars in F2 to K1 spectral range should have very significant chemical compositions differences.

    The number and locations of spectral lines varies with the surface effective temperature. The effect of differences in effective temperature should therefore give a dependence of the

number of detections with spectral type even within the F2 to K1 spectral range. One would expect the strength and locations of the lines, and therefore the strength of the Fourier peak, to increase gradually with decreasing effective temperature, reach a maximum then gradually decrease with decreasing temperatures. Consequently one would expect a triangle-like distribution in Figure 16. The box-like distribution seen in Figure 16 that normalizes the number of detected stars to the number of detected F9 stars, using the relative distribution of spectral types in the SEGUE 2 survey, shows that there is no correlation of peak strength with spectral type. It is highly peculiar that the strength and the number of lines do not vary within the F2 to K1 spectral range. Furthermore the stars in the halo come from early generations of stars formation and are therefore expected to have fewer spectral lines than later generations, making it less likely that they have the required large number of peculiar lines. Note also that 1 of the detected stars in Figure 16 and 2 of the detected stars in Figure 12 have been classified as A0. It is highly unlikely that some A stars would have such a high number of strong peculiar lines.

      Although there is no physical reason why a gravity effect can generate the signal, one may wonder whether it might. The discussion in section 5.2 considers the only spectral features (spectral lines located at nearly periodic frequency locations) that could generate the signal, concluding that too many spectral lines located at nearly periodic locations would be needed. There is no physical reason why giants and supergiants would have these peculiar spectral lines. Figures 11 and 13 show that the majority of the stars with a detected signal are F stars. The majority of stars in the SDSS and SEGUE surveys are in the halo of our galaxy, consequently the F stars that contain the signal, cannot be giants and supergiants, since F giants and supergiants would have to be relatively young stars, while the stars in the halo are old stars. Yanny et al. (2009) discuss the presence of giants in the SEGUE SDSS survey. They clearly state that giants, that would have a spectral type earlier than G, are not present in the SEGUE survey. On the other hand, the SEGUE survey targeted K giants ([Yanny](Yanny) et al. 2009) so that a significant fraction of K stars in the SEGUE 2 survey are giants. If we assume that the signals are present in giants, we should have detected a large number of signals in K stars. However, Figure 16, that shows the number of detected stars in the SEGUE 2 survey normalized to the number of detected F9 stars, does not show an excess of detections in K

stars. To the contrary, there is a detection of K1 stars equal to the detection of F9 stars, only a small number of detections of K3 stars and no detections at all for K5 and K7 stars.

Differences in rotational velocity or turbulence introduce a Doppler Effect broadening that changes the width of all of the spectral lines with a widening that vary slowly with frequency and wavelength. To begin, one must consider that the SDSS spectra have low spectral resolutions $(\lambda/\Delta\lambda \sim 2000)$. Using $\Delta\lambda/\lambda = 1/2000 = v/c$ this spectral resolution gives a velocity $v = 150$ Km/sec. As shown in Figures 12, 14 and 16, the overwhelming majority of the stars that have a detected signal have a spectral type ranging between F9 and K1 where the rotational velocities are very small. Turbulence causes a small Doppler Effect. Consequently the low SDSS spectral resolution dominates over turbulent or rotational broadening effects in the F9 to K1 spectral range and makes them irrelevant.

Radial velocity shifts the location of the spectral lines so that different radial velocities cause different shifts. This can have an effect because of the discrete sampling that comes from the pixels of the CCD that detects the spectra. This effect can be relevant because of the low resolution of the SDSS spectra gives a small number of samplings (3900) in the frequency spectra analyzed and therefore large sampling intervals. Because these sampling intervals are larger than the widths of the spectral lines of stars in the F9 and K1 spectral range, different radial velocities can cause a weakening or strengthening of the observed lines that depends on the location of the lines and the radial velocity of the stars. Different radial velocities of the stars should therefore affect a signal that comes from the Fourier transform of the lines. We shall now consider the weakening caused by the Doppler Effect from radial velocities to explain why only 1% of the solar type stars have a signal. Computer simulations were carried out to quantitatively estimate the effect of radial velocities. We used the SDSS spectra of G2 stars that have the $N = 764$ signal, Doppler shifted the spectrum and then took the Fourier transform of the spectrum to evaluate the effect on the strength of the Fourier signal. We found that a radial velocity shift of 150 Km/sec reduces the strength of the peak by a factor of 1.15 and that a shift of 300 Km/sec decreases the strength by a factor of 1.4. The dashed line at the top of Figure 25 shows that a weakening of the signal by a factor of 1.5 predicts a strong increase of the fraction of detections with *SNR* that is in total contradiction with the observations given by the black dots. Consequently, the majority of the stars in the SDSS would need a radial velocity much larger than 300 Km/sec to induce the required weakening.

This is totally impossible since the SDSS stars are in the halo of our galaxy and the distribution of the number of halo stars as a function of radial velocity has a dispersion of 100 Km/sec. Consequently the overwhelming majority of the SDSS stars have radial velocities smaller than 300 Km/sec. Only a very small fraction of the galactic stars, called hypervelocity stars, have velocities higher than 300 km/sec. The average rotational velocity of the Earth around the sun is only 30 Km/sec and therefore will have a negligible effect. The stars that have detected signals in Table 1 and 2 have radial velocities ranging between -150 and +120 Km/sec, in agreement with a random sampling of the distribution of the number of halo stars as a function of radial velocity. One may wonder why the computer simulations gave a weakening of the Fourier signal. This is because the radial velocities also affect the signal generated by the Fourier transform of a periodic spectral modulation. The $N = 765$ signal gives a cosine modulation in the frequency spectra with a period of $6.127 \; 10^{11}$ Hz that is only a factor of 6 larger than the sampling resolution of the SDSS spectra. Consequently some change with radial velocity should be expected. To intuitively understand this effect, consider a hypothetical case where the sampling is exactly equal to half of the period of a cosine spectral modulation. If the redshift is zero, the periodic variation is clearly detected because if a sampling is at the maximum intensity value then the next sampling is at the minimum intensity value and consequently the periodical signal, given by the intensity difference, is detected with the highest strength. Let us now consider the effect of a radial velocity that generates a shift that varies with frequency: The samplings no longer occur at the best locations and, consequently, the strength of the signal decreases because the intensity differences decrease.

We can therefore conclude that effects coming differences in spectral lines strength, chemical composition, effective temperature, gravity, rotational velocities, turbulence or radial velocities cannot be responsible for the small fraction of detections in the F9 to K1 spectral range.

7. **Possible physical causes**

*7.1 Rapid Pulsations*

In principle, rapid pulsations in small regions of the atmospheres of the stars could generate the signals (Borra 2010). However, the periods in Figure 2 show that the period of the pulsation would have to be of the order of $1.65 \; 10^{-12}$ seconds, which appears unrealistically small for stars.

*7.2 Signal Generated by Extraterrestrial Intelligence*

We shall now consider the possibility that the signals are generated by Extraterrestrial Intelligence (ETI), following the suggestion in Borra (2012). Borra (2012) uses computer simulations to create a spectrum having a modulation, periodic in frequency units, generated by pairs of pulses separated by a constant value of time ($\Delta t = 10^{-14}$ seconds). Figure 5 in Borra (2012) shows a computer generated frequency spectrum that contains such a signal and figure 6 in Borra (2012) shows the Fourier transform of this spectrum. Like in our case, the signal is not visible in the noisy frequency spectrum (figure 5 in Borra 2012), while it is clearly visible in its Fourier spectrum (figure 6 in Borra 2012). The detected signals therefore have exactly the characteristics of the signals predicted by Borra (2012). In particular, the fact that the detected Fourier signals have a very narrow width, within the sampling limit of the FFT, indicates that the spectral modulation is present in most of the spectral range of the SDSS spectrum and we can assume that it even goes beyond. Borra (2012) also shows spectra that have a strong spectral modulation visible in the spectra themselves (e.g. figure 4 in Borra (2012)) and one may wonder why ETI does not send such a strong signal, particularly if one considers that ETI is very likely to have a far more advanced technology than we are and consequently to have extremely powerful sources (Borra 2012). There may be several reasons. Firstly sending a stronger signal requires more energy, which has obvious inconveniences, particularly if one considers signals sent to a very large number of stars. Furthermore ETI obviously knows that Fourier transforms exist and can detect weak signals, like we did. Finally, ETI knows that very technologically advanced civilizations can have powerful telescopes that can therefore detect weak signals. We can understand this last statement by considering how telescopes have evolved on Earth over the last century and will certainly improve considerably over the next centuries. In the more distant future there is the possibility of having extremely powerful telescopes. For example Angel et al. (2008) consider the possibility of a 100-m diameter mirror on the moon.

In section 4 we state that the signals contain a few times $10^{-5}$ of the energy in the spectrum. Prima facie, this seems like a very large value since one would intuitively assume that it is a few times $10^{-5}$ of the total energy emitted by the star. However, this is only the fraction detected by us. ETI would probably send it in a very small beam aimed at particular stars so that the energy actually emitted would be a considerably smaller fraction of $10^{-5}$ of the energy of the star. For example Howard et al. (2004), who discuss searches of nanosecond optical pulses from nearby solar stars generated by ETI, consider signals generated with a laser and emitted in a beam having a 20 milliarcseconds diameter. Section 3 in Borra (2012) discusses in more details the physical considerations of an ETI signal.

The analysis at the end of section 4, that discusses instrumental and data reduction effects, estimates that the fraction of signals detected is consistent with a signal coming from approximately 1 % of the stars in the F2 to K1 spectral range. This analysis makes the assumption that all the spectra have a signal having the same intensity. This assumption is valid in the case of instrumental and data reduction effects but not in the case of signal originating from ETI. The effect of signals having varying intensities would simply add noise to the continuous line in Figure 22. In the case of the ETI hypothesis we would expect ETI to generate a signal significantly above the contribution of the stellar background, within the narrow beam, to facilitate its detection. This is the kind of signal that generated the continuous line in Figure 22, which fits well the data, where we assume that there is a signal significantly stronger than the signal used in computing the continuous in Figure 21, which does not fit the data.

The fact that only a small fraction of stars, in a narrow spectral range centered near the spectral type of the sun, has the signal is in agreement with the ETI hypothesis since we intuitively would expect that solar type stars would have ETI and only a small fraction of them would have ETI.

8. **Conclusion**

After carrying out a Fourier analysis of 2.5 million spectra in the SDSS survey, we have found signals indicating a periodic modulation in the spectra of a very small number of stars that are all within a small spectral range centered near the spectral type of the sun. It was not found in any of the spectra of quasars and galaxies. We considered several possible effects that could cause the signals.

Instrumental or data analysis effects could cause the spectral modulation. However, this hypothesis is not valid because of three main reasons (see section 4): Firstly, a signal generated by instrumental or data analysis effects should be present in the spectra of all the quasars stars and galaxies, while it is present in only a very small fraction of F and G stars and none in galaxies. Secondly, signals should mostly be detected in the brightest stars because of signal to noise ratio considerations, while we find that this is not the case. Thirdly, and most importantly, the periodic spectral modulation signal was only detected in a very small number of stars that are contained within a narrow range of spectral types (F2 to K1). As discussed in section 4 the probability that this is due to random fluctuations is $< 10^{-16}$ for the results from the SEGUE 2 data and is also very small for the results from the Data Release 8 SDSS data. Finally, we find that the detections do not occur preferentially in some fibers nor at particular Julian Dates as might be the case for some type of instrumental effects. The detections occur within large time spans and therefore cannot be due to occasional instrumental problems (e.g. caused by instrumental changes). Note also that we cannot invoke a hypothetical instrumental color effect to explain the fact that all the detections are within a narrow spectral range, because the signal was not detected in any of the galaxies that have spectra with similar mean signal to noise ratios, while most galaxies have colors within that spectral range. We cannot assume that this lack of detection is due to the extendedness of the galaxies because the SDSS spectra are obtained with optical fibers that have a diameter of 3 arcseconds, comparable to the broadening caused by atmospheric seeing in stars.

The Fourier transform of spectral features is another possible source of the Fourier signal. However a quantitative analysis, that uses computer simulations, shows that that it is highly unlikely because several hundreds of strong lines placed at nearly periodic frequency locations would be needed. The fact that, as discussed in section 5.2, the strengths of the signals are contained within a relatively small range of strengths and are present in only 1 % of the stars in a narrow spectral range also makes it unlikely since one would expect the lines to be present in all the stars in a given spectral type (e.g. G2) and have a wide range of strengths in the F2 to K1 spectral range.

We considered effects that might change the shapes and locations of the spectral lines and therefore the strength and location of the signal coming from the Fourier transform of the lines. Assuming that the signals come from the Fourier transform of spectral lines, these strength and location variations might be responsible for the small fraction of detections given by the black dots in figures 21, 22 and 25. The effects could come from differences in chemical composition, effective temperature, gravity, rotational velocities, turbulence and radial velocities. The modelling carried out in section 6 shows that a hypothetical effect would have to cause a signal weakening that increases very rapidly with the increase of the effect. We find that radial velocity is the only relevant effect, since computer simulations show that increasing radial velocities decreases the strength of the signal. However radial velocities considerably larger than 300 km/sec in 99% of the stars in the halo of the galaxy are needed to give a significant effect. This cannot be the case because the overwhelming majority of the stars come from the SDSS SEGUE 1 and SEGUE 2 survey which targeted faint stars in the stellar halo of the galaxy. The overwhelming majority of these stars have radial velocities smaller than 300 Km/sec.

We considered the possibility that the spectra of molecules generate the signal, concluding that the hypothesis does not hold.

Finally, we considered the possibility that the signals are caused by intensity pulses generated by Extraterrestrial Intelligence (ETI), as suggested by Borra (2012), to make us aware of their existence. The shape of the detected signals has exactly the shape predicted by Borra (2012). The ETI hypothesis is strengthened by the fact that the signals are found in stars having spectral types within a narrow spectral range centered near the G2 spectral type of the sun. Intuitively, we would expect stars having a spectral type similar to the sun to be more likely to have planets capable of having ETI. This is a complex and highly speculative issue (see Lammer et al. 2009) and we shall not delve on it. Let us however note that all of the active optical SETI observational projects listed in Tarter (2001) search for signals in Solar-type stars. Reines & Marcy (2002) and Howard et al. (2004) searched for signals generated by lasers in nearby solar stars. In particular, let us note that Howard et al. (2004) searched for nanosecond optical pulses from nearby solar stars.

The ETI hypothesis requires that all different ETI transmitters choose to broadcast with the same time separation of pulses and one may wonder why they do so. This is a highly speculative issue that may have several explanations. A possible explanation that makes sense is that all ETI use the same time separation to make it clear that the pulses all come from ETI.

At first sight, one may question the validity of the ETI hypothesis because of the energy required to send the pulses to distant stars. The energy issue is discussed in Borra (2012) that shows that technology presently available on Earth could be used to send signals having the energy needed to be detected 1000 light years away. Obviously, more advanced civilizations would have technologies capable to generate much stronger signals; Borra (2012) elaborates on this. As an illustration of this, just imagine how the suggestion made by Borra (2012) would have been considered if submitted in 1950, before the invention of the laser, when it would have suggested the use of a light bulb to send the signal.

At this stage, the ETI generation of the spectral modulation is a hypothesis that needs to be confirmed with further work. This can be done by repeatedly observing the stars in Tables 1 and 2 with photoelectric detectors capable of detecting very rapid intensity signals. However ETI may not necessarily send us pulses at all times so that a lack of detections in some stars may not necessarily signify that ETI does not exist. The reason why ETI may not send pulses at all times may simply come from the fact that the signals must be sent to a very large number of stars so that too much energy would be required to send pulses to all stars at all times. The kind of signal that we have detected can be generated by pairs of pulses that have the same time separation $\Delta t$ but with the pairs sent with time separations significantly larger than $\Delta t$ (Borra 2012). One could therefore look for the ETI pulses using techniques similar to the one described in Leeb et al. (2013) because pairs of pulses separated by a constant value of $\Delta t = 1.6465\ 10^{-13}$ seconds could be sent with a periodicity having a period much larger than $\Delta t$ (e.g. $10^{-6}$ seconds). Leeb et al. (2013) estimate that a telescope having a 1.7-m diameter could detect signals from a G2V star 500 *ly* distant so that this type of signal could be detected in stars to distances as large as 2000 *ly* with existing telescopes. However the detected stars in Tables 1 and 2 are at distances greater than 8000 *ly*, so that a 30-m telescope would be needed to observe the stars listed in tables 1 and 2. One could however carry out Fourier transforms on spectra of nearby stars to search for periodic signals and then observe detected stars with the technique of Leeb et al. (2013) and smaller telescopes. Finally, note that the larger time separations between the pairs of pulses separated by the same $\Delta t = 1.6465\ 10^{-13}$ seconds do not have to be periodic. The pairs could be sent at random or with time separations modulated at will, provided the time separation $\Delta t$ between the pairs remains constant. Consequently ETI may add more information in the signals by sending pairs of pulses separated by the same constant time ( $\Delta t = 1{,}6465 \; 10^{-13}$ seconds ) but with the pairs sent in a Morse-like code to send us messages or perhaps even pictures of themselves.

The objects listed in Tables 1 and 2 should also be observed with large telescopes to obtain spectra with high resolutions and high signal to noise ratios that would allow studying the signals in greater details to definitely confirm that they are not data reduction or instrumental effects.

## Acknowledgements

This research has been supported by the Natural Sciences and Engineering Research Council of Canada. Funding for SDSS-III has been provided by the Alfred P. Sloan Foundation, the Participating Institutions, the National Science Foundation, and the U.S. Department of Energy Office of Science. The SDSS-III web site is http://www.sdss3.org/.

**References**

Angel, R., Worden, S.P. Borra, E.F., Eisenstein, D. J. et al. 2008, ApJ, 680, 1582

Borra, E.F. 2010, A&A letters, 511, L6

Borra, E.F. 2012, AJ, 144, 181

Borra, E.F. 2013, ApJ, 774.142

Chin, S.L., Francois, V., Watson, J.M., & Delisle, C. 1992, Applied Optics, 31, 3383

Cocconi, G., & Morrison, P. A. 1959, Nature, 184, 844

Grevesse, N. & Sauval, J. 1994, Molecules in the sun and molecular data: Molecules in the Stellar Environment. Springer Berlin Heidelberg, 1994. 196

Howard, A. W, Horowitz, P., Wilkinson, D. T., et al. 2004, ApJ, 613, 1270

Korpela, E. J., Anderson, D. P., Bankay, R., et al. 2011, Proc. SPIE, 8152, 815212

Lammer, H., Bredehoft, J.H., Coustenis, A., et al. Astron, Astrophys. Rev. 17, 181

Leeb, W.R. , Poppe, 1 A. , Hammel, E. , Alves, J. Brunner, M. & Meingast, S. 2013, Astrobiology, 13, 521

Reines, A. E., & Marcy, G. W. 2002 , PASP, 114,416

Rothman, L.S. et al. 2009, Journal of Quantitative Spectroscopy and Radiative Transfer, 110, 533

Sinha , K. 1991. Proc Astronomical society of Australia, 9, 32

Tarter, J. 2001, Ann. Rev . Astron. & Astrophys., 39, 611

Trottier, E. 2012, Recherche de signaux périodiques dans des spectres astronomiques, M.Sc. Thesis, Université Laval

Yanny, B. et al. 2009, AJ 137, 4377

FIGURES

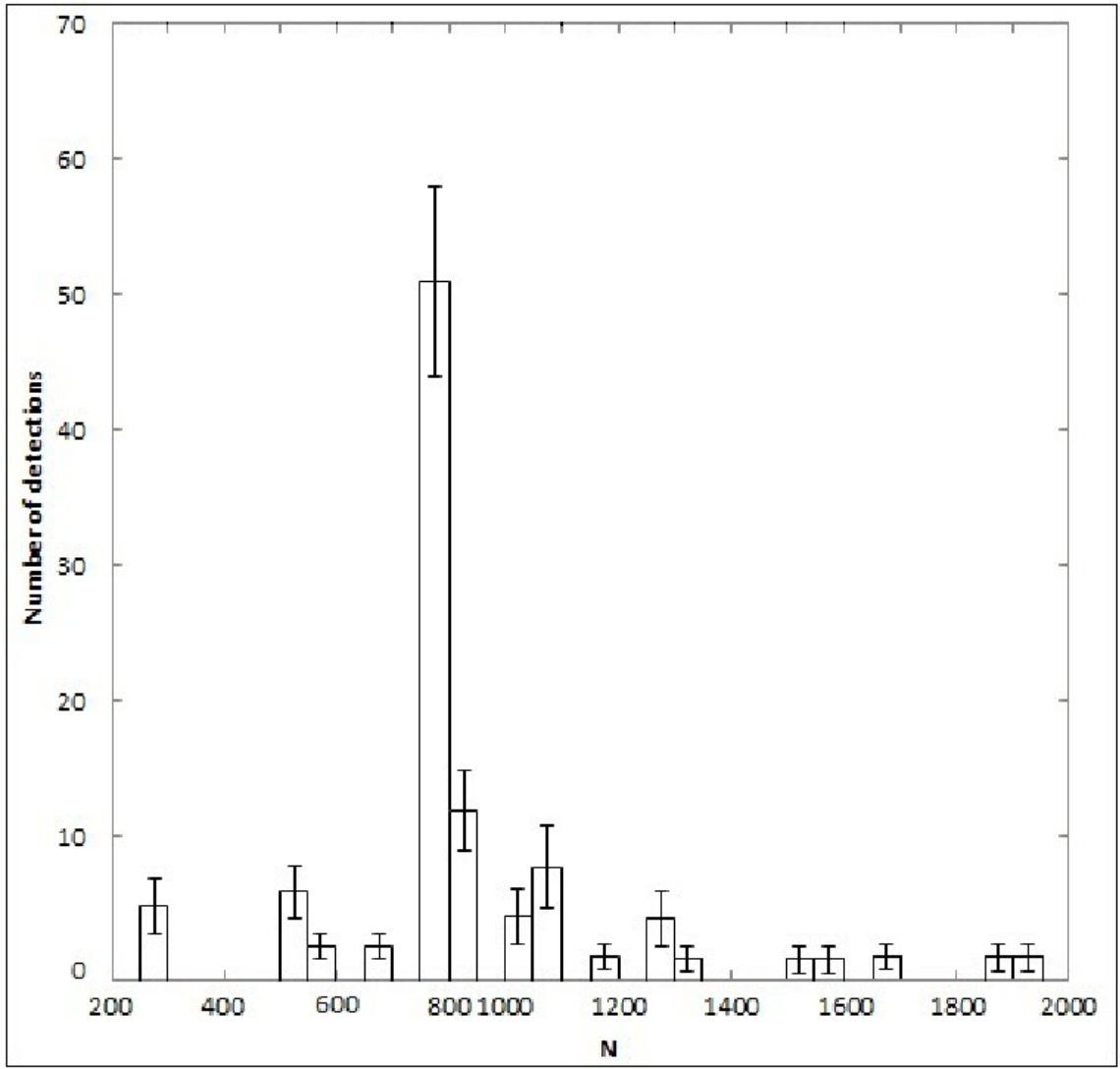

Figure 1

Distribution of detected signals (for stars only) as a function of the sampling number $N$ at which they are detected in the Fourier spectrum. To convert to time units (seconds) $N$ must be multiplied by $2.1\ 10^{-15}$. The error bars show +- a standard deviation. Note that 45 out of a total of 51 of the detected signals in the $750 < N < 800$ box are located at $N = 764$ or $N = 765$ and, as discussed in section 3 and seen in Figure 2, are actually located at the same $N$ location after an appropriate correction.

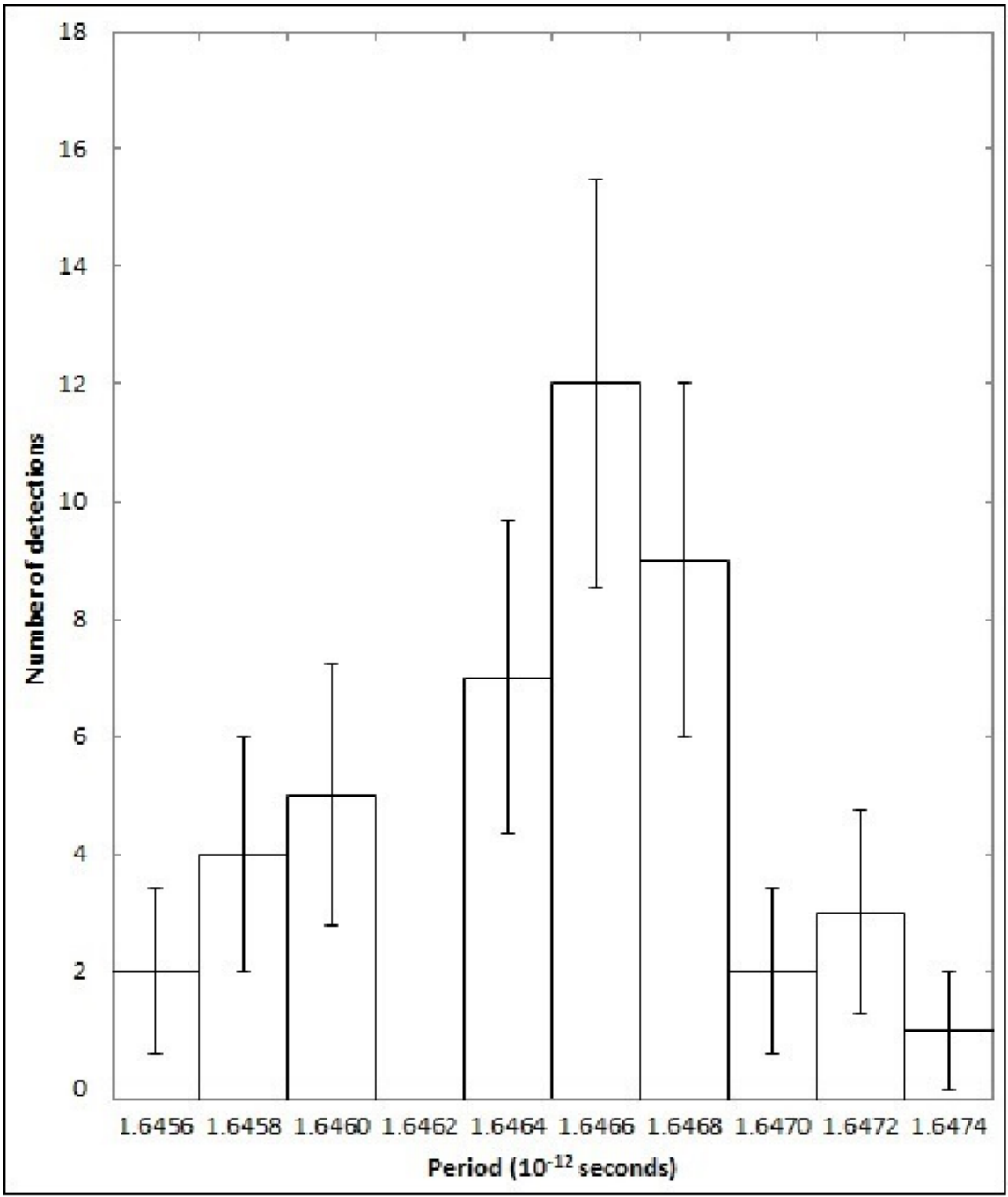

Figure 2

Location of the majority of the signals present in the strongest box in Figure 1 (750 < *N* < 800 ) after conversion to time units. The total width of the horizontal axis is within the width of the sampling limit (ΔN = 1), which indicates that the spectral modulation covers the entire spectral range of the SDSS spectrum. The error bars show +- a standard deviation.

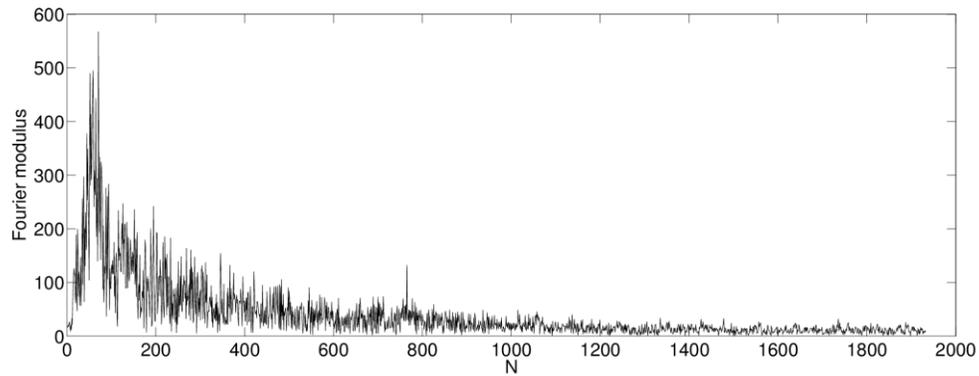

Figure 3

Fourier modulus of the frequency spectrum (after subtraction of its smoothed spectrum) of an F5 star that has a statistically significant signal. The Fourier modulus is plotted as a function of the FFT sampling number N.

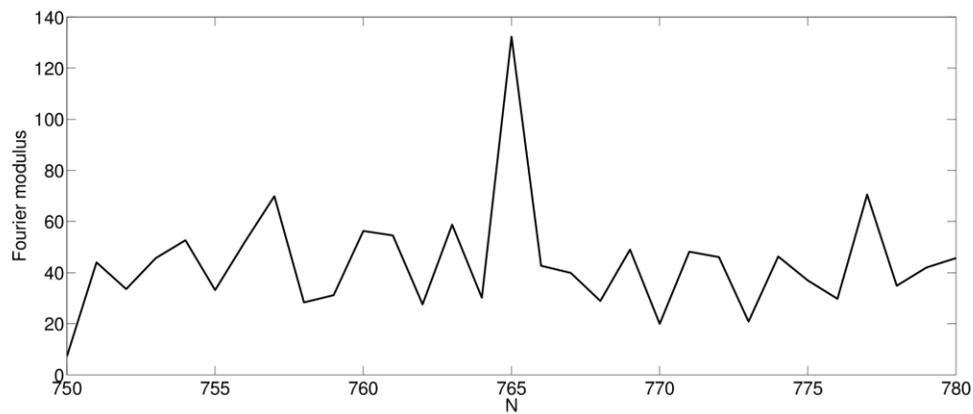

Figure 4

Zoom in the region of the signal for the Fourier spectrum plotted in Figure 3.

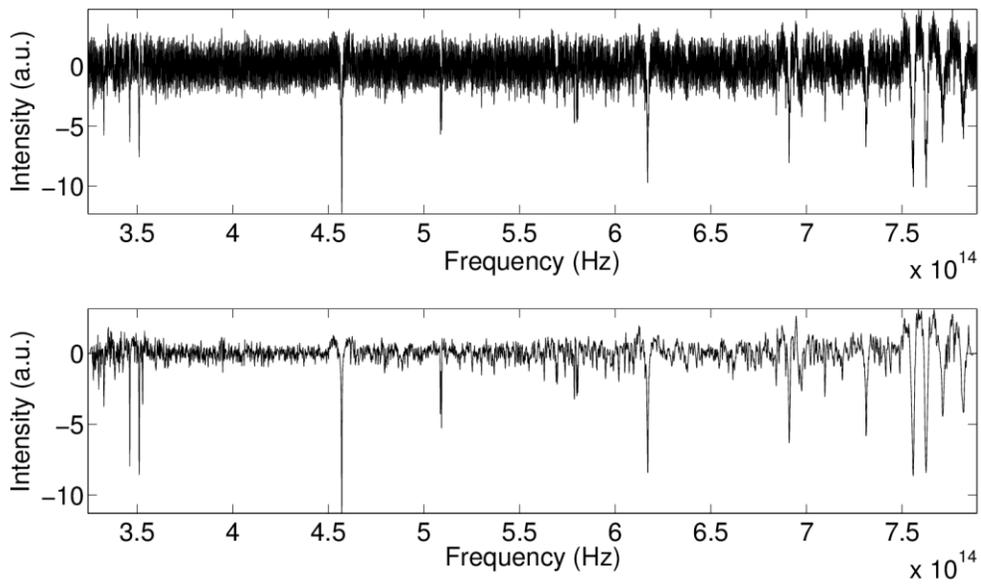

Figure 5

The bottom plot shows the unaltered frequency spectrum of the star that generated Figures 3 and 4. The intensities are in arbitrary units (a.u.). The sinusoidal signal is totally undetectable by eye inspection because it has very small amplitude. To make the characteristics of the signal visible, we added a signal that has the period of the detected signal but with amplitude multiplied by 10 ^4. This spectrum is plotted at the top of the figure.

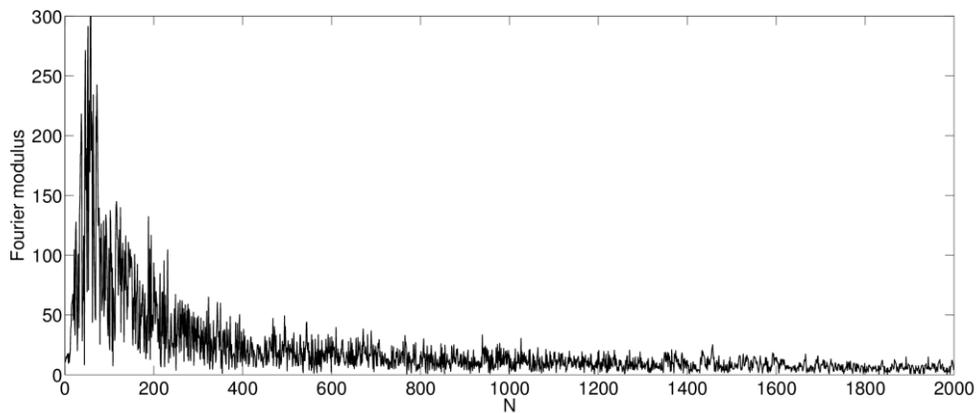

Figure 6

Fourier modulus of the frequency spectrum (after subtraction of its smoothed spectrum) of an F star template taken from the SDSS database. The Fourier modulus is plotted as a function of the FFT sampling number $N$.

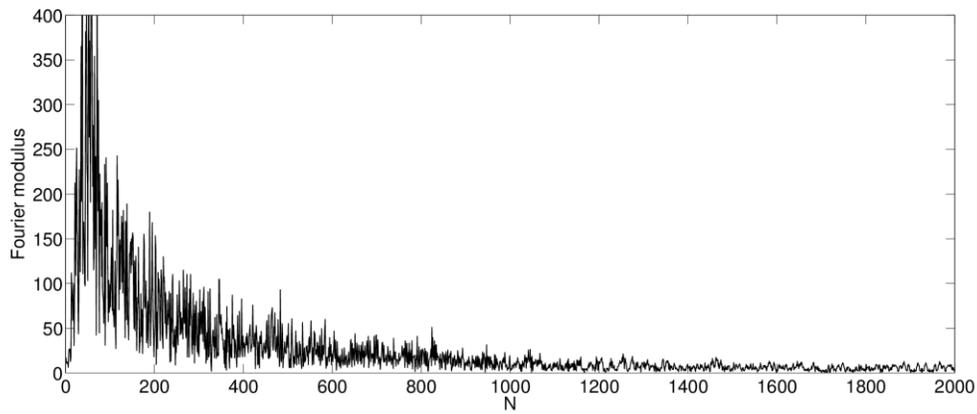

Figure 7

Fourier modulus of the frequency spectrum (after subtraction of its smoothed spectrum) of an F5 star that did not have a signal. The Fourier modulus is plotted as a function of the FFT sampling number $N$.

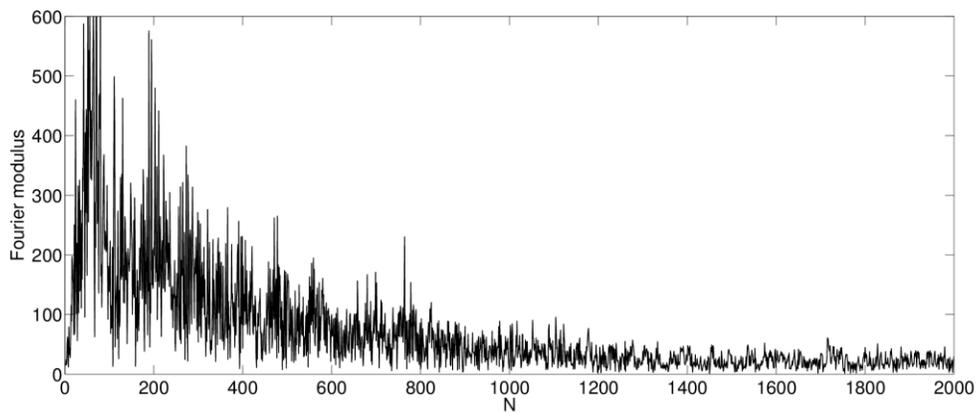

Figure 8

Fourier modulus of the frequency spectrum (after subtraction of its smoothed spectrum) of a K1 star that has a statistically significant signal. The Fourier modulus is plotted as a function of the FFT sampling number *N*.

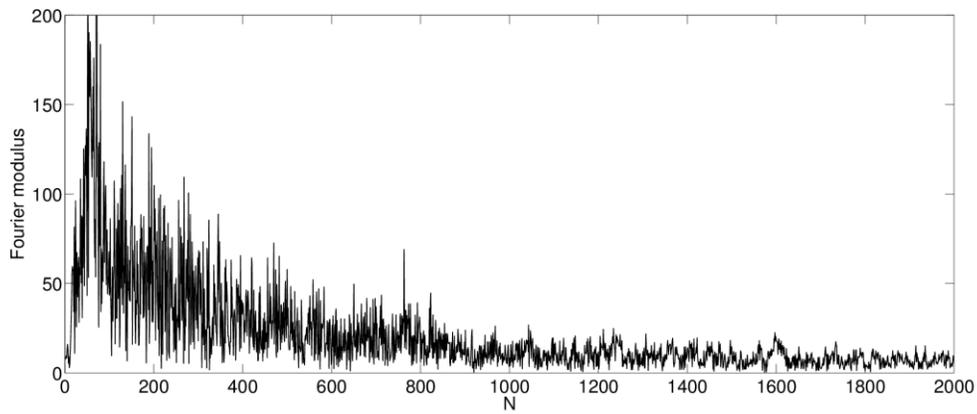

Figure 9

Fourier modulus of the frequency spectrum (after subtraction of its smoothed spectrum) of a G2 star that has a statistically significant signal. The Fourier modulus is plotted as a function of the FFT sampling number *N*.

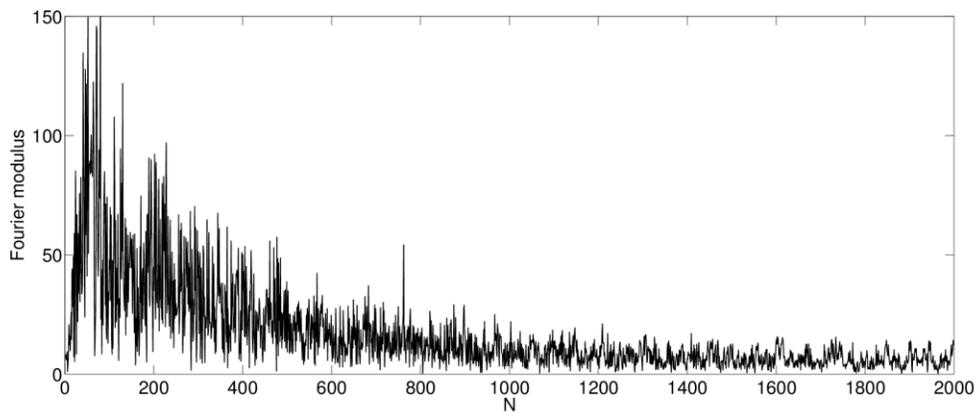

Figure 10

Fourier modulus of the frequency spectrum (after subtraction of its smoothed spectrum) of another F5star that has a statistically significant signal. The Fourier modulus is plotted as a function of the FFT sampling number *N*.

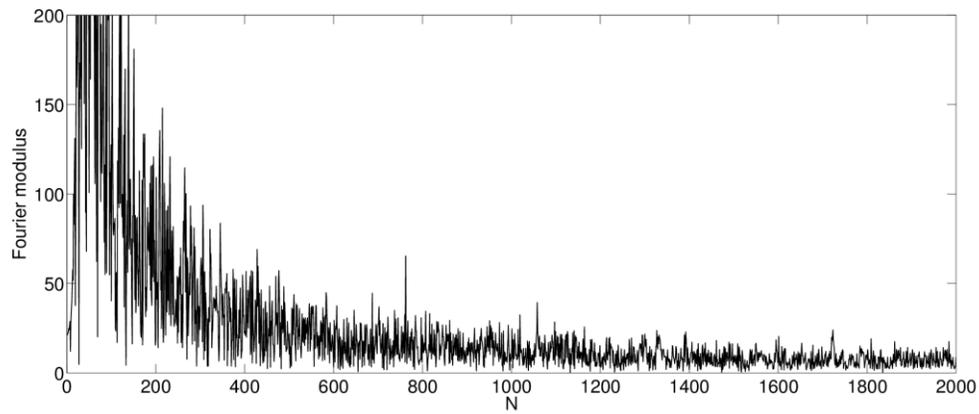

Figure 11

Fourier modulus of the frequency spectrum (after subtraction of its smoothed spectrum) of an A0 star that has a statistically significant signal. The Fourier modulus is plotted as a function of the FFT sampling number *N*.

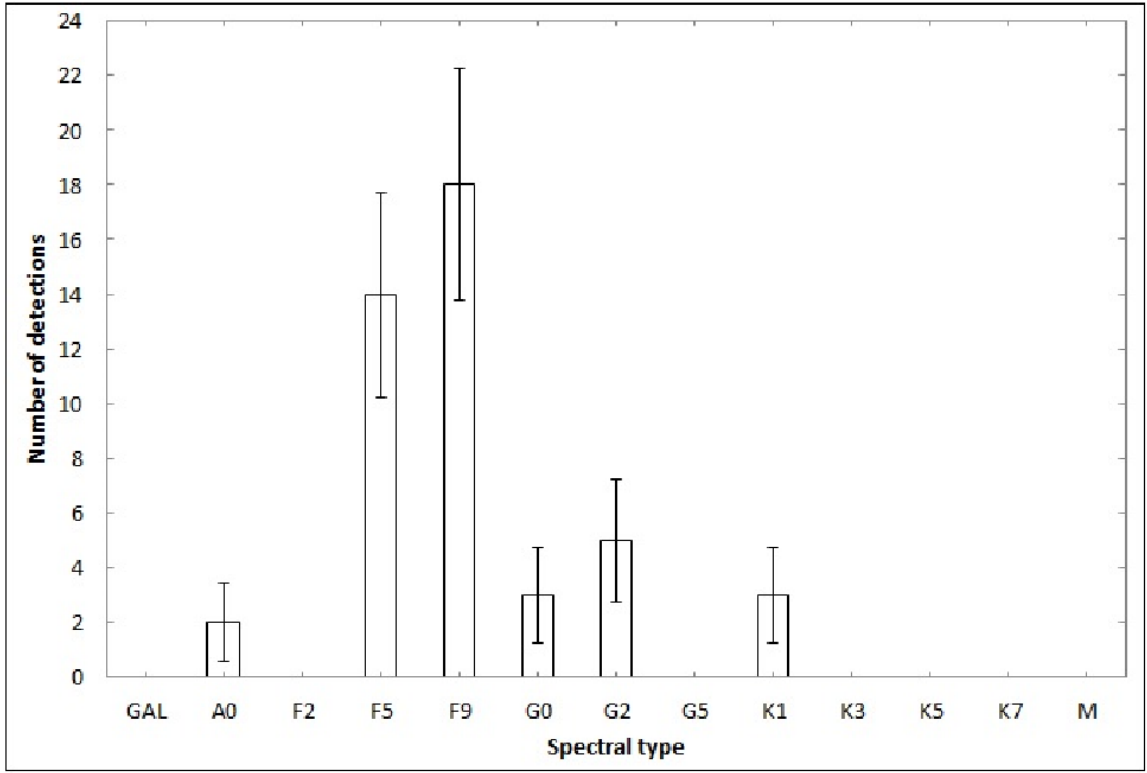

Figure 12

Histogram of the distribution of the number of stars that have a detected signal within the time range in Figure 2 as a function of spectral type. The error bars show +- a standard deviation.

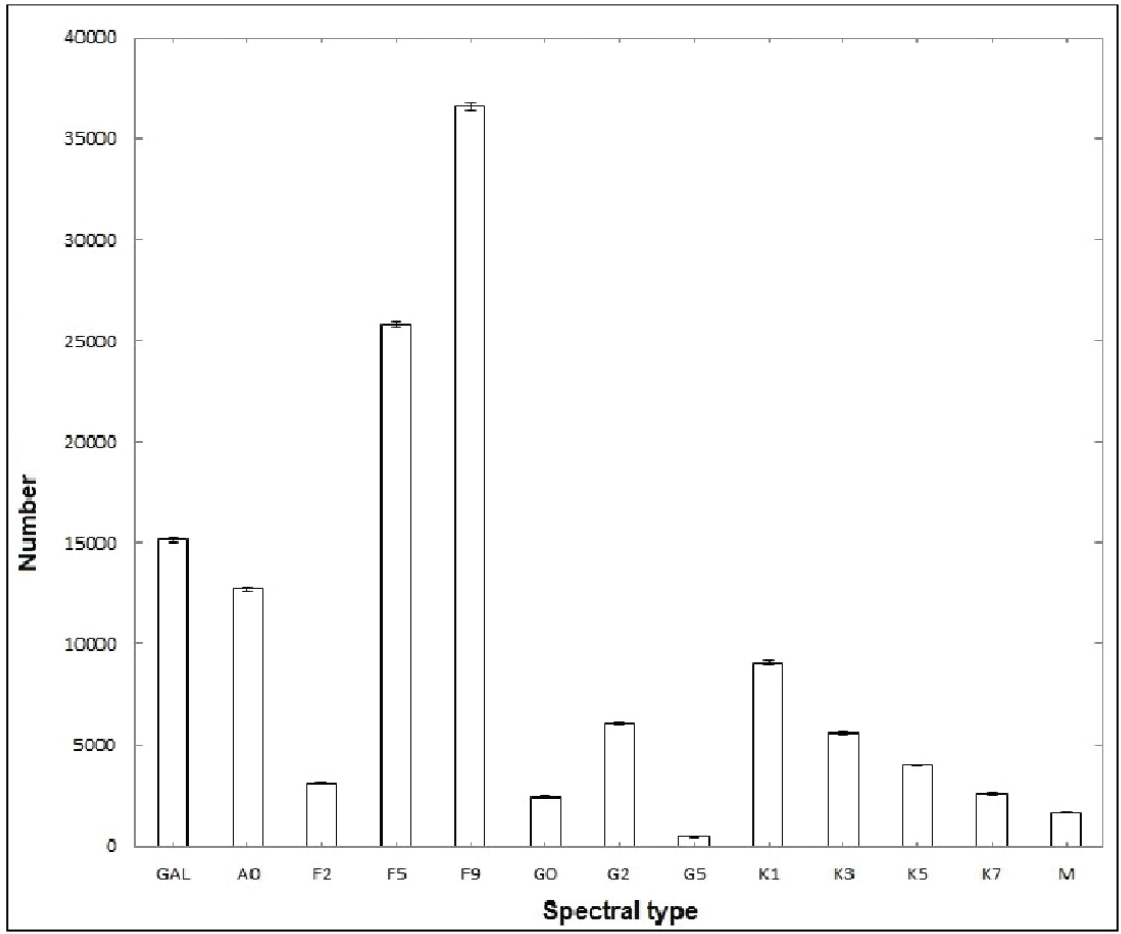

Figure 13

Histogram of the distribution of all the spectral types that are in the SDSS data analyzed, including galaxies. For a proper comparison with Figure 12, only stars and galaxies having a signal to noise ratio comparable to the stars in Figure 12 are included. The error bars show +- a standard deviation and have very small amplitudes in the figure.

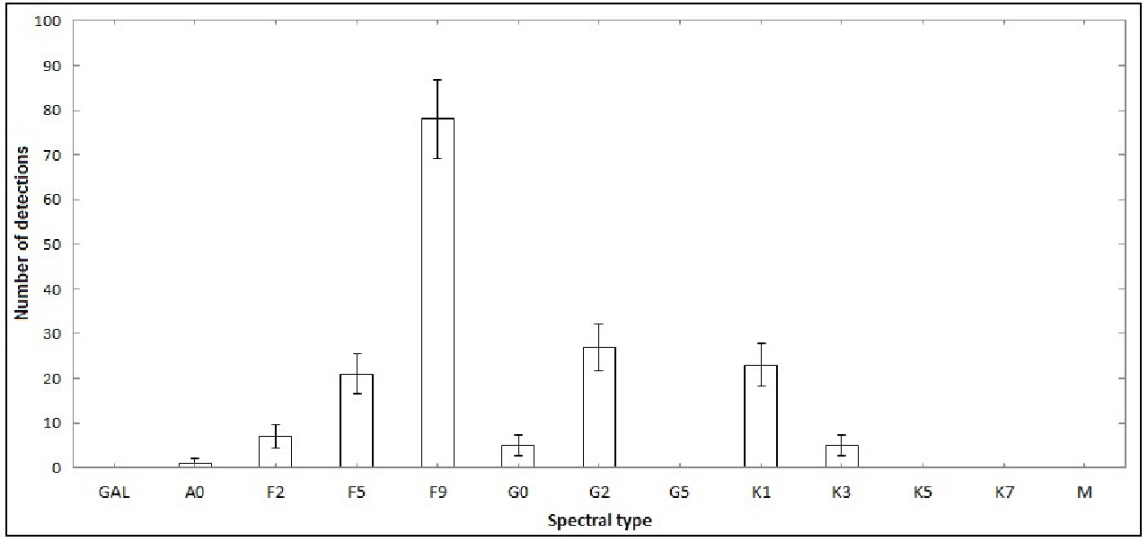

Figure 14

Histogram of the distribution of the detected signals that are within the time range in Figure 2 as a function of spectral type in the SEGUE 2 survey. Like in Figure 12 the majority of the signals are detected in spectral types ranging from F5 to K1. The error bars show +- a standard deviation.

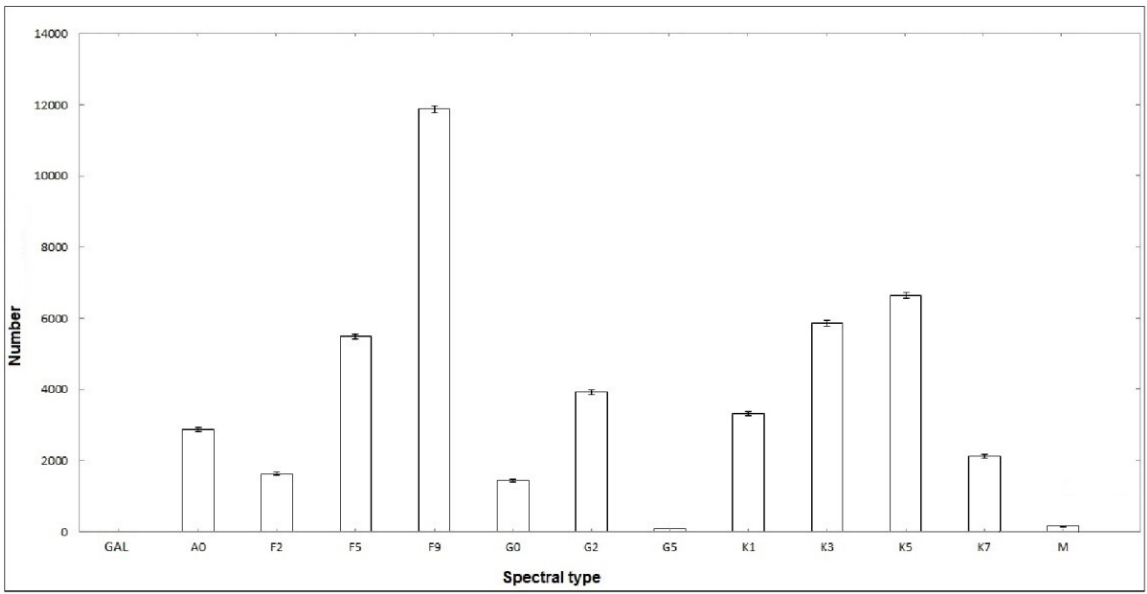

Figure 15

Histogram of all the spectral types in the SEGUE 2 survey. For a proper comparison with Figure 14, only stars having a signal to noise ratio comparable to the stars in Figure 14 are included. The error bars show +- a standard deviation and have very small amplitudes in the figure.

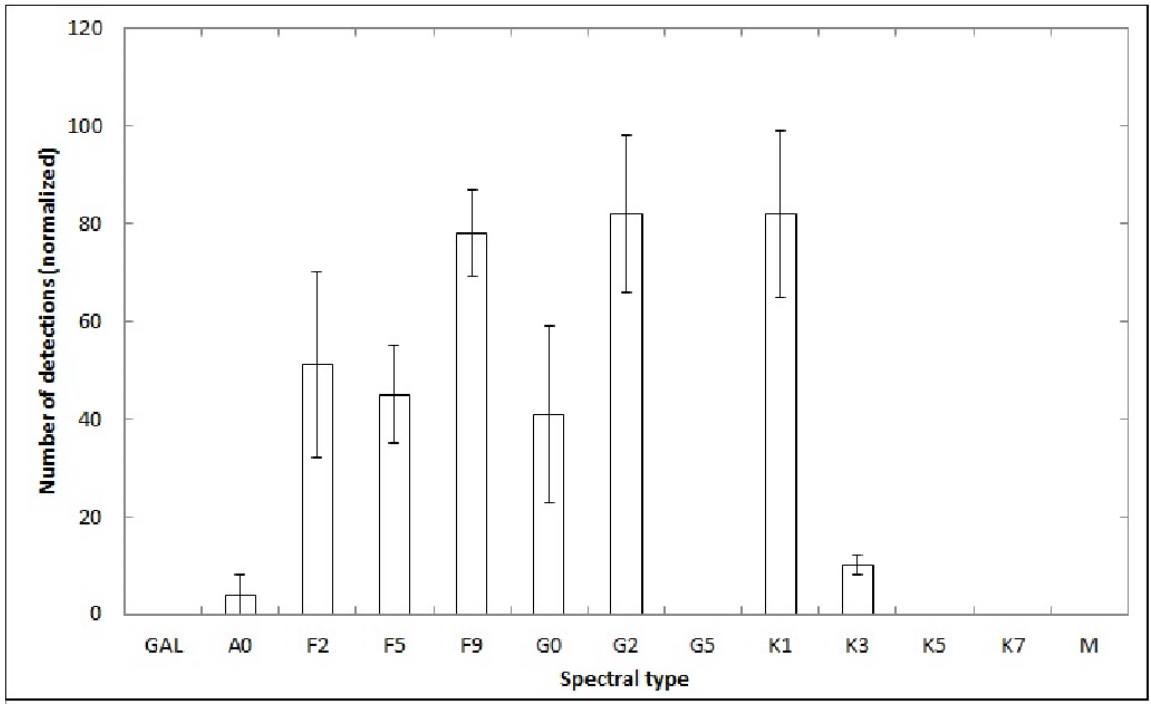

Figure 16

Number of detected stars in the SEGUE 2 survey normalized to the number of detected F9 stars, using the relative distribution of spectral types in Figure 15. The error bars show +- a standard deviation.

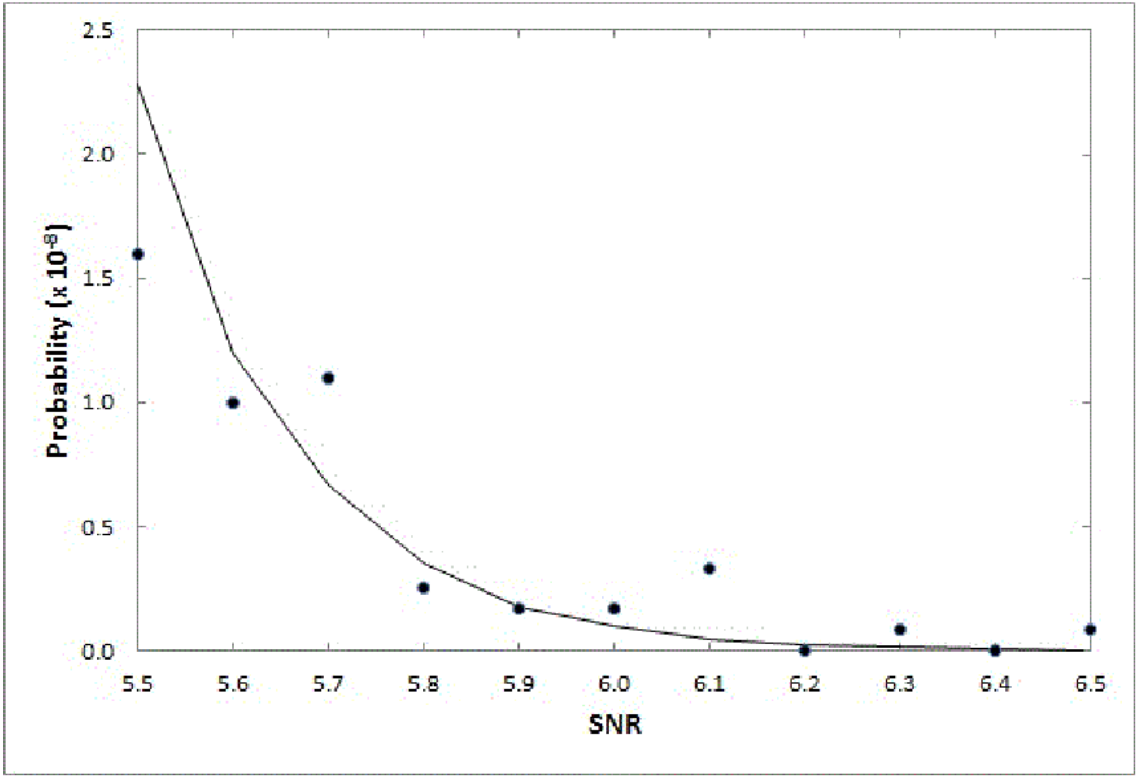

Figure 17

The continuous line gives the theoretical Rayleigh distribution used, while the filled circles give the fraction of the number of detections of peaks in stars. It shows a good agreement between the theoretical and observed numbers. We do not include the stars with peaks detected in the range $762 < N < 765$, because these are peaks in the narrow range where we have found real signals.

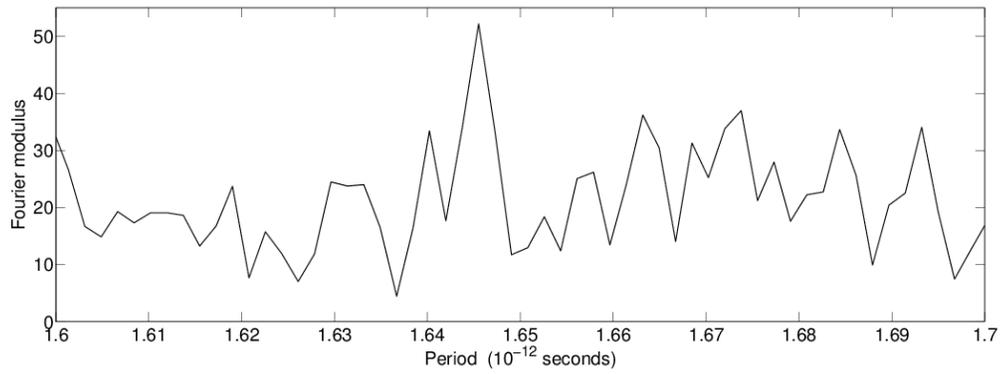

Figure 18

Fourier transform of the blue region of the same spectrum that was used to generate Figures 3 and 4.

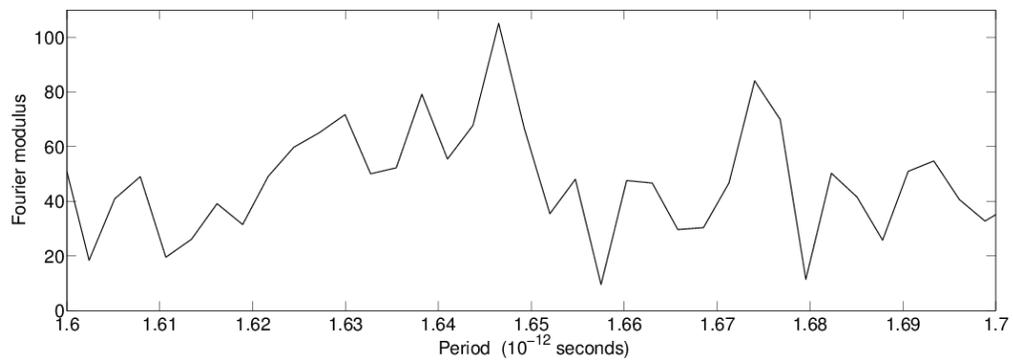

Figure 19

Fourier transform of the red region of the same spectrum that was used to generate Figures 3 and 4.

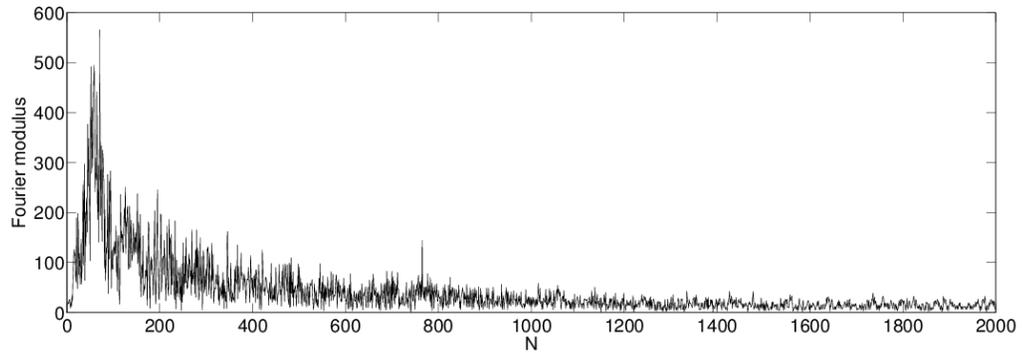

Figure 20

Fourier transform of the spectrum used to generate Figure 3 but obtained with a Fourier Transform that does not use a linear interpolation.

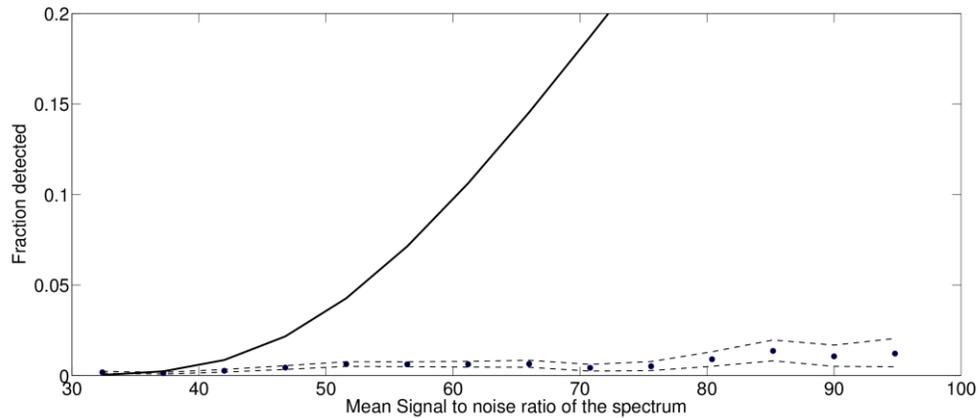

Figure 21

The black dots, at the very bottom of the figure, give the fraction of detected stars, obtained by dividing the number of detected stars by the number of observed stars in a spectral range from F2 to K1, as a function of the median signal to noise ratio of the spectrum. The two dashed lines show the region surrounding the black dots within +/- a standard deviation. The continuous line shows the increase of the fraction of detected stars predicted by Rayleigh statistics if we assume that the signal is present in all stars within a spectral range from F2 to K1.

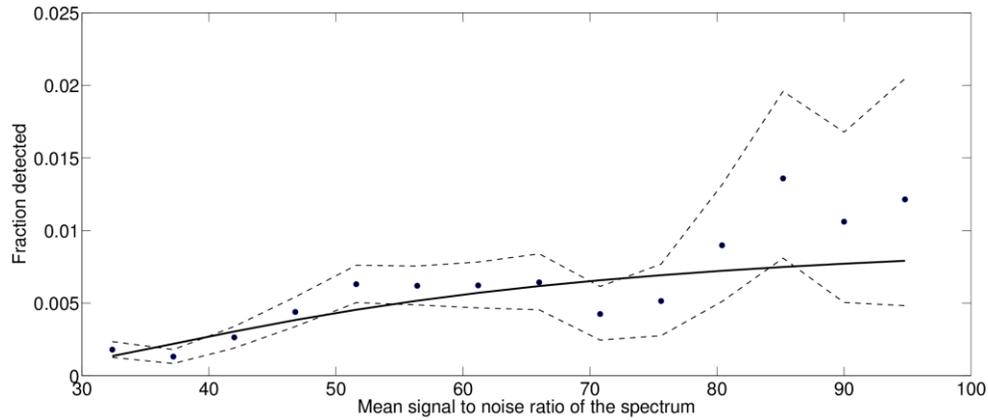

Figure 22

Like in Figure 21, the black dots give the fraction of detected stars, obtained by dividing the number of detected stars by the number of observed stars within a spectral range from F2 to K1, as a function of the median signal to noise ratio of the spectrum. The two dashed lines show the region surrounding the black dots within +/- a standard deviation. The continuous line shows the increase of the fraction of detected stars predicted by Rayleigh statistics if we assume that there is a signal, significantly stronger than the signal assumed in Figure 21 that is present in only 1% of the stars within a spectral range from F2 to K1.

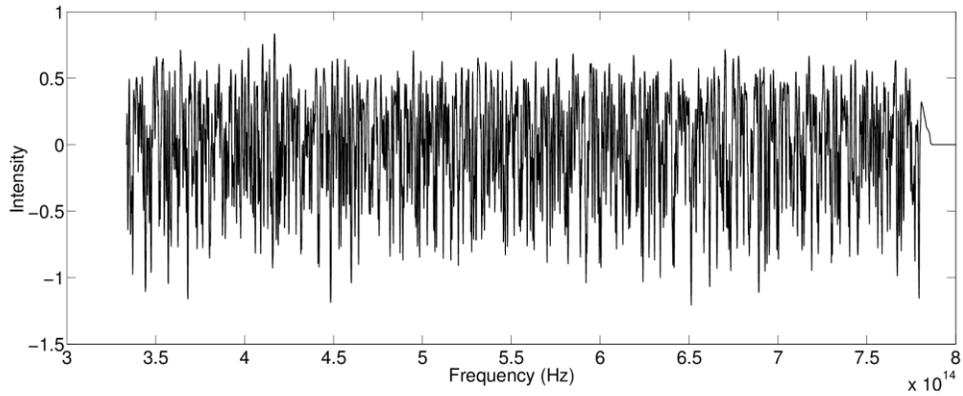

Figure 23

Spectrum of a computer simulation which starts from a Shah model with 729 lines separated by 6.127 $10^{11}$ Hz. The intensity of the lines and their central locations are varied at random with the technique described in section 4.

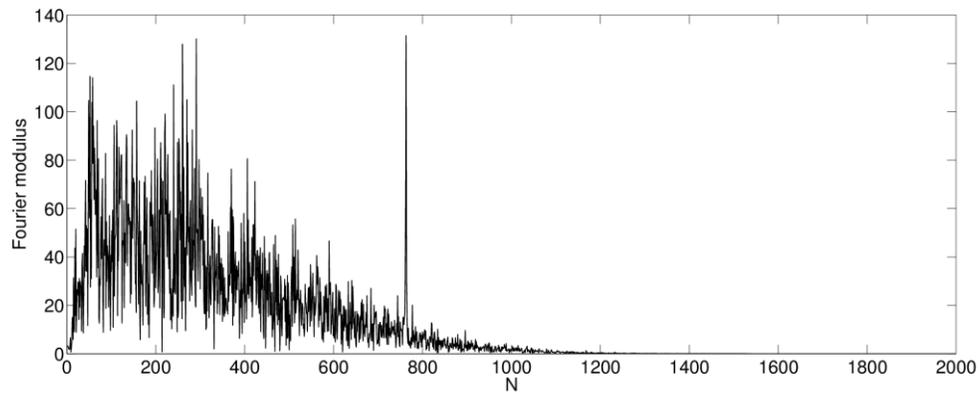

Figure 24

Fourier transform of the computer simulated spectrum in Figure 23.

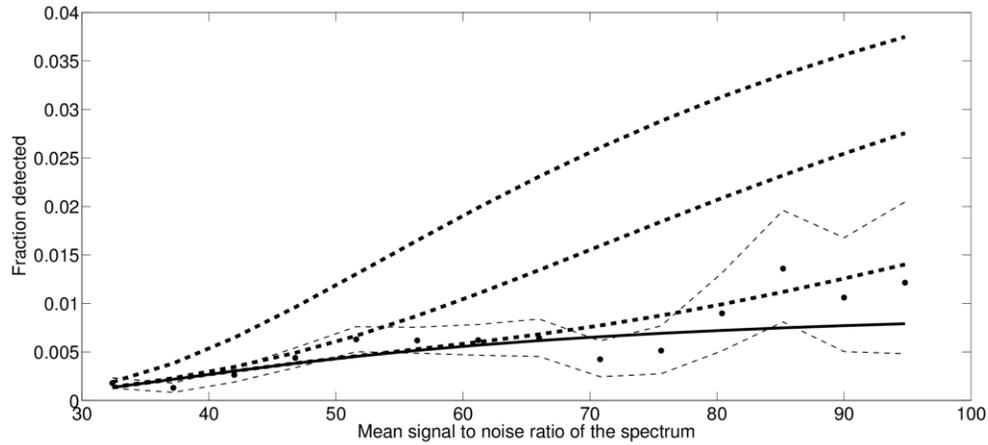

Figure 25

The dashed lines show the results of the modelling discussed in this section. The top dashed line shows the results in the case where the signal is weakened by a factor of 1.5, the middle dashed line shows the results for a signal weakened by a factor of 2 and the bottom dashed line shows the results for a signal weakened by a factor of 3.

Table 1
Stars detected in the SDSS Data Release 8 survey

| Plate ID | Fiber | MJD (-2400000) | RA | Dec | Spectrum | g | SNR |
|---|---|---|---|---|---|---|---|
| 1135 | 355 | 53024 | 7.56566 | 0.04684 | F5 | 16.89 | 5.84 |
| 2336 | 227 | 53712 | 20.40206 | 39.30142 | G2 | 17.00 | 5.55 |
| 2043 | 248 | 53351 | 20.81054 | 38.49003 | F9 | 17.13 | 5.68 |
| 2850 | 190 | 54461 | 25.13918 | -9.74885 | F9 | 16.37 | 5.77 |
| 2045 | 247 | 53350 | 29.31848 | -0.98506 | F9 | 16.44 | 5.80 |
| 2378 | 451 | 53759 | 43.73998 | 35.40452 | F5 | 17.12 | 5.59 |
| 1512 | 537 | 53035 | 44.4278 | 0.03721 | F5 | 15.82 | 5.82 |
| 811 | 404 | 52669 | 47.3618 | 0.97422 | F9 | 16.94 | 5.66 |
| 2334 | 291 | 53730 | 50.42148 | 4.28652 | F5 | 15.69 | 6.09 |
| 2714 | 387 | 54208 | 117.25872 | 18.54427 | F5 | 15.83 | 5.68 |
| 2549 | 243 | 54523 | 122.27698 | 82.16647 | K1 | 19.76 | 5.53 |
| 2057 | 211 | 53816 | 123.26652 | -0.49372 | F5 | 15.15 | 5.78 |
| 2315 | 122 | 53741 | 127.76086 | 22.9905 | F9 | 16.05 | 5.71 |
| 2315 | 452 | 53741 | 127.89986 | 25.64138 | F9 | 17.92 | 5.56 |
| 2383 | 376 | 53800 | 145.69073 | 62.99296 | K1 | 16.5 | 5.61 |
| 558 | 284 | 52317 | 148.07319 | 56.78558 | K1 | 16.88 | 6.00 |
| 1236 | 270 | 52751 | 150.08079 | 7.80431 | G2 | 0.00 | 5.65 |
| 270 | 359 | 51909 | 151.13819 | 0.05867 | F9 | 15.58 | 5.60 |
| 2106 | 468 | 53714 | 166.10635 | 32.93721 | F9 | 17.15 | 5.56 |
| 2856 | 573 | 54463 | 167.58321 | 39.22128 | F9 | 16.38 | 5.51 |
| 2856 | 58 | 54463 | 167.73098 | 37.62736 | F9 | 16.88 | 5.77 |
| 2394 | 437 | 54551 | 168.32244 | 59.03954 | F5 | 16.22 | 5.56 |
| 280 | 258 | 51612 | 170.13999 | -0.57195 | F5 | 16.21 | 5.57 |
| 2861 | 315 | 54583 | 170.98471 | -7.9588 | G2 | 16.05 | 5.94 |
| 1996 | 19 | 53436 | 172.55219 | 38.90159 | F9 | 16.39 | 5.65 |
| 2862 | 192 | 54471 | 173.80539 | -0.3558 | G0 | 16.62 | 5.52 |
| 2467 | 38 | 54176 | 184.13554 | 39.58293 | F5 | 16.94 | 5.52 |
| 2895 | 356 | 54567 | 187.64251 | 0.21593 | F9 | 17.64 | 5.58 |
| 2925 | 435 | 54584 | 197.15766 | 39.48877 | F5 | 17.06 | 5.65 |
| 2925 | 487 | 54584 | 197.7972 | 39.6416 | F5 | 16.79 | 5.86 |
| 2617 | 62 | 54502 | 197.86139 | 18.9493 | G0 | 15.89 | 5.56 |
| 2929 | 274 | 54616 | 205.64524 | 27.53287 | G2 | 16.9 | 5.53 |
| 2929 | 432 | 54616 | 206.67425 | 28.95404 | F5 | 16.96 | 5.62 |
| 2931 | 527 | 54590 | 212.73705 | 37.72277 | F9 | 18.23 | 5.51 |
| 1380 | 243 | 53084 | 213.80491 | 37.7377 | A0 | 15.23 | 5.95 |
| 2447 | 564 | 54498 | 216.05652 | 57.0435 | F9 | 16.67 | 5.63 |
| 2909 | 380 | 54653 | 221.259 | 1.02268 | F9 | 17.93 | 5.51 |
| 2766 | 150 | 54242 | 228.58601 | 14.87639 | A0 | 16.01 | 5.62 |
| 1167 | 159 | 52738 | 235.33978 | 48.37447 | G0 | 17.45 | 5.80 |
| 2174 | 609 | 53521 | 251.63594 | 36.51416 | F9 | 17.3 | 5.75 |
| 2561 | 114 | 54597 | 263.79883 | 63.75807 | F5 | 16.57 | 5.62 |
| 2552 | 360 | 54632 | 275.5926 | 64.21574 | F9 | 15.92 | 5.51 |
| 2800 | 624 | 54326 | 291.85221 | 38.27539 | F5 | 15.85 | 5.51 |
| 2553 | 574 | 54631 | 292.24331 | 63.38305 | G2 | 16.5 | 5.61 |
| 1475 | 182 | 52903 | 331.3208 | -0.19869 | F9 | 18.02 | 5.72 |

| Plate ID | Fiber | MJD (-2400000) | RA | Dec | Spectrum | g | SNR |
|---|---|---|---|---|---|---|---|
| 1093 | 243 | 52591 | 354.76279 | -1.19339 | F9 | 16.49 | 5.92 |

Table 2
Stars detected in the SEGUE 2 survey

| Plate ID | Fiber | MJD (-2400000) | RA | Dec | Spectrum | g | SNR |
|---|---|---|---|---|---|---|---|
| 3134 | 51 | 54806 | 2.56492 | -7.821 | G2 | 15.91 | 5.05 |
| 3133 | 342 | 54789 | 7.80968 | 15.35036 | F9 | 16.73 | 5.44 |
| 3133 | 326 | 54789 | 7.97119 | 15.82425 | F9 | 16.14 | 5.19 |
| 3133 | 372 | 54789 | 8.56401 | 15.70373 | F9 | 16.46 | 5.60 |
| 3133 | 363 | 54789 | 8.62795 | 15.99092 | F5 | 15.71 | 5.20 |
| 3133 | 550 | 54789 | 9.6991 | 15.48404 | F9 | 16.92 | 5.00 |
| 3133 | 584 | 54789 | 10.27366 | 15.02305 | F9 | 16.89 | 5.00 |
| 3111 | 381 | 54800 | 12.37016 | 0.38197 | F5 | 15.75 | 5.02 |
| 3111 | 445 | 54800 | 13.3716 | 0.84059 | F2 | 18.2 | 5.09 |
| 3112 | 196 | 54802 | 14.15 | 0.17395 | F9 | 16.2 | 5.07 |
| 3112 | 569 | 54802 | 15.6438 | 0.80556 | A0 | 15.87 | 5.05 |
| 3114 | 402 | 54773 | 28.5233 | 14.9999 | F9 | 17.29 | 5.18 |
| 3122 | 535 | 54821 | 35.58029 | -7.66537 | F9 | 17.41 | 5.03 |
| 3127 | 292 | 54835 | 35.62846 | -0.61817 | F5 | 16.44 | 5.54 |
| 3241 | 112 | 54884 | 37.71358 | 23.38642 | K1 | 16.57 | 5.04 |
| 3241 | 69 | 54884 | 37.88569 | 23.3171 | G2 | 16.19 | 5.17 |
| 3126 | 535 | 54804 | 39.88496 | -0.01177 | F9 | 16.23 | 5.02 |
| 3210 | 590 | 54876 | 43.57155 | 32.78828 | F5 | 18.89 | 5.02 |
| 3186 | 272 | 54833 | 48.09157 | -7.08649 | K3 | 17.06 | 5.29 |
| 3183 | 569 | 54833 | 49.82055 | 1.23683 | G0 | 16.85 | 5.90 |
| 3186 | 590 | 54833 | 50.06875 | -6.73085 | F5 | 16.26 | 5.04 |
| 3187 | 303 | 54821 | 50.37162 | 17.42007 | F5 | 16.45 | 5.19 |
| 3187 | 530 | 54821 | 52.26521 | 18.6307 | F9 | 17.78 | 5.16 |
| 3187 | 580 | 54821 | 52.46462 | 18.65872 | K1 | 17.07 | 5.02 |
| 3156 | 598 | 54792 | 56.08844 | 0.7961 | F9 | 16.84 | 5.08 |
| 3123 | 579 | 54741 | 73.8798 | -3.74702 | K1 | 17.99 | 5.17 |
| 3209 | 392 | 54906 | 74.53448 | -3.55246 | F5 | 16.19 | 5.08 |
| 3334 | 210 | 54927 | 103.00474 | 15.6352 | F5 | 15.24 | 5.08 |
| 3207 | 269 | 54850 | 103.03763 | 29.03584 | F5 | 16.2 | 5.02 |
| 3207 | 352 | 54850 | 103.06319 | 29.65777 | G2 | 17.03 | 5.32 |
| 3335 | 127 | 54922 | 103.07852 | 16.35379 | F5 | 16.40 | 5.30 |
| 3335 | 162 | 54922 | 103.27128 | 16.49979 | F5 | 16.41 | 5.34 |
| 3207 | 30 | 54850 | 105.54896 | 28.6141 | K1 | 16.89 | 5.24 |
| 3206 | 89 | 54852 | 106.12668 | 28.54312 | F9 | 17.5 | 5.03 |
| 3206 | 42 | 54852 | 106.64924 | 28.83633 | K1 | 18.41 | 5.01 |
| 3206 | 544 | 54852 | 106.90776 | 30.12719 | K1 | 18.44 | 5.11 |
| 3205 | 270 | 54848 | 108.42185 | 39.74213 | F2 | 16.11 | 5.36 |
| 3225 | 71 | 54853 | 115.78622 | 43.22074 | F9 | 16.59 | 5.48 |
| 3227 | 347 | 54893 | 116.55394 | 27.97063 | F9 | 18.18 | 5.11 |
| 3161 | 62 | 54779 | 117.21568 | 40.81777 | F2 | 17.27 | 5.12 |
| 3161 | 548 | 54779 | 117.35744 | 41.50262 | F2 | 17.04 | 5.03 |
| 3227 | 495 | 54893 | 118.29786 | 27.70467 | G0 | 16.87 | 5.11 |
| 3208 | 270 | 54853 | 118.41663 | 66.20397 | K1 | 16.99 | 5.20 |
| 3227 | 589 | 54893 | 119.07456 | 27.97372 | K1 | 17.06 | 5.31 |
| 3227 | 594 | 54877 | 119.18344 | 27.60045 | K3 | 18.38 | 5.17 |
| 3227 | 611 | 54877 | 119.46281 | 28.32816 | F9 | 16.76 | 5.12 |
| 3227 | 590 | 54877 | 119.47617 | 27.63328 | G2 | 17.14 | 5.18 |
| 3226 | 312 | 54857 | 121.15657 | 18.49633 | F9 | 16.96 | 5.01 |
| 3226 | 52 | 54857 | 123.2541 | 17.79068 | F9 | 16.94 | 5.02 |
| 3188 | 352 | 54831 | 123.31434 | 31.13033 | K3 | 17.13 | 5.18 |
| 3226 | 26 | 54857 | 123.99834 | 18.10461 | F9 | 17.99 | 5.16 |
| 3153 | 237 | 54790 | 125.12921 | 39.71875 | F9 | 16.5 | 5.21 |
| 3188 | 566 | 54831 | 125.42759 | 31.89675 | K3 | 17.65 | 5.11 |
| 3188 | 573 | 54831 | 125.72641 | 31.64771 | F9 | 17.38 | 5.06 |
| 3182 | 246 | 54828 | 126.10674 | 31.33025 | F9 | 17.76 | 5.45 |
| 3229 | 151 | 54876 | 126.46452 | 37.26601 | F5 | 16.38 | 5.23 |
| 3228 | 551 | 54863 | 126.67435 | 33.06847 | F9 | 16.87 | 5.43 |
| 3175 | 311 | 54828 | 127.58891 | 46.86221 | K1 | 16.61 | 5.14 |
| 3153 | 596 | 54790 | 127.65426 | 39.84967 | F5 | 15.85 | 5.42 |
| 3182 | 29 | 54828 | 128.53438 | 31.30428 | G2 | 16.81 | 5.06 |
| 3175 | 252 | 54828 | 128.76172 | 46.62887 | F9 | 16.87 | 5.15 |
| 3175 | 170 | 54828 | 129.62226 | 46.65653 | F9 | 17.06 | 5.36 |
| 3293 | 88 | 54921 | 130.46436 | 4.62557 | F9 | 17.38 | 5.12 |
| 3293 | 91 | 54921 | 130.48466 | 4.78351 | G2 | 16.13 | 5.09 |
| 3293 | 48 | 54921 | 130.55417 | 4.47424 | F9 | 17.94 | 5.11 |
| 3293 | 613 | 54921 | 131.32869 | 6.51254 | K3 | 18.58 | 5.04 |
| 3293 | 628 | 54921 | 131.68078 | 5.92109 | F9 | 17.96 | 5.04 |
| 3149 | 92 | 54806 | 136.79845 | 37.27433 | F9 | 16.35 | 5.06 |
| 3149 | 144 | 54806 | 136.98931 | 37.89869 | F9 | 17.48 | 5.05 |
| 3264 | 72 | 54889 | 140.89188 | 43.30854 | F5 | 16.95 | 5.02 |
| 3192 | 48 | 54829 | 141.53814 | 13.66854 | F9 | 17.06 | 5.13 |
| 3317 | 170 | 54908 | 141.56794 | 20.05715 | K1 | 16.41 | 5.11 |
| 3319 | 252 | 54915 | 141.64173 | 9.02133 | F9 | 16.6 | 5.05 |
| 3317 | 490 | 54908 | 141.88661 | 20.60915 | K1 | 16.93 | 5.09 |
| 3212 | 452 | 54851 | 142.13383 | 24.1452 | F9 | 16.68 | 5.38 |
| 3196 | 272 | 54834 | 142.34652 | 14.64321 | F9 | 16.27 | 5.09 |
| 3196 | 234 | 54834 | 142.46288 | 14.85168 | G2 | 17.66 | 5.2 |
| 3196 | 209 | 54834 | 143.00361 | 13.93461 | F5 | 15.99 | 5.04 |
| 3196 | 190 | 54834 | 143.05308 | 14.25917 | G2 | 17.06 | 5.32 |
| 3196 | 49 | 54834 | 144.05754 | 14.1358 | F9 | 16.7 | 5.08 |
| 3151 | 606 | 54804 | 144.68692 | 8.15839 | K1 | 17.19 | 5.12 |
| 3223 | 71 | 54865 | 146.80052 | 36.2722 | K1 | 16.49 | 5.3 |
| 3223 | 10 | 54865 | 146.94098 | 35.85537 | F9 | 16.5 | 5.16 |
| 3321 | 534 | 54924 | 147.23791 | 44.60509 | F9 | 17.78 | 5.08 |
| 3320 | 573 | 54912 | 149.82984 | 52.13923 | F9 | 17.34 | 5.03 |
| 3154 | 573 | 54821 | 150.57821 | 2.9022 | F9 | 17.45 | 5.09 |
| 3257 | 315 | 54888 | 150.67385 | 2.62065 | K1 | 18.03 | 5.82 |
| 3194 | 607 | 54833 | 151.83539 | 43.01633 | F9 | 17.27 | 5.05 |
| 3179 | 372 | 54830 | 151.91836 | 21.67181 | F9 | 17.35 | 5.17 |
| 3257 | 412 | 54888 | 151.96443 | 4.03443 | G2 | 16.24 | 5.13 |
| 3179 | 211 | 54830 | 152.19297 | 19.78518 | F5 | 16.63 | 5.05 |
| 3178 | 270 | 54848 | 152.7232 | 18.40982 | F9 | 16.26 | 5.23 |
| 3178 | 392 | 54848 | 152.79772 | 19.30112 | F2 | 15.92 | 5.09 |
| 3287 | 352 | 54941 | 153.05452 | 43.00246 | F9 | 16.37 | 5.03 |
| 3324 | 472 | 54943 | 153.24285 | 54.21501 | G2 | 17.17 | 5.34 |
| 3178 | 412 | 54848 | 153.41605 | 19.59436 | G2 | 15.86 | 5.15 |
| 3178 | 170 | 54848 | 153.70075 | 18.22239 | G2 | 15.88 | 5.07 |
| 3179 | 622 | 54830 | 153.97324 | 21.34077 | G2 | 16.44 | 5.04 |
| 3177 | 312 | 54833 | 154.09452 | 55.55876 | F9 | 15.82 | 5.23 |
| 3260 | 538 | 54883 | 154.45088 | 28.99442 | K1 | 18.61 | 5.19 |
| 3178 | 632 | 54848 | 155.1668 | 19.02867 | F9 | 16.53 | 5.09 |
| 3250 | 429 | 54883 | 159.30435 | 24.31519 | K1 | 16.69 | 5.14 |
| 3299 | 151 | 54908 | 161.95254 | 15.68697 | F9 | 15.88 | 5.11 |
| 3000 | 71 | 54892 | 166.08248 | 40.17829 | F9 | 11.4 | 5.18 |
| 3000 | 37 | 54843 | 166.09362 | 40.28256 | F9 | 14.01 | 5.07 |
| 3243 | 571 | 54910 | 167.24277 | 3.99046 | F9 | 17.3 | 5.22 |
| 3326 | 71 | 54943 | 168.59182 | 40.86181 | F9 | 16.63 | 5.45 |
| 3216 | 289 | 54853 | 169.06288 | 45.56936 | F9 | 17.08 | 5.15 |
| 3327 | 367 | 54951 | 169.96724 | 18.34179 | F9 | 18.14 | 5.04 |
| 3327 | 197 | 54951 | 170.03549 | 16.90242 | F9 | 18.22 | 5.04 |
| 3170 | 374 | 54907 | 171.04333 | 20.37785 | G2 | 17.16 | 5.26 |
| 3233 | 252 | 54891 | 171.57288 | -2.19868 | F9 | 15.7 | 5.04 |
| 3233 | 407 | 54891 | 171.86536 | 0.2547 | F9 | 17.52 | 5.48 |
| 3233 | 467 | 54891 | 171.97609 | -0.74836 | K1 | 17.89 | 5.09 |
| 3233 | 552 | 54891 | 172.74336 | -0.73158 | F9 | 16.91 | 5.01 |
| 3233 | 32 | 54891 | 173.17079 | -1.35412 | F9 | 16.13 | 5.3 |
| 3244 | 289 | 54892 | 173.5424 | 12.44855 | F9 | 17.13 | 5.13 |
| 3244 | 428 | 54892 | 174.01085 | 13.98754 | K1 | 18.14 | 5.05 |
| 3221 | 423 | 54864 | 174.54703 | 26.96446 | F5 | 18.96 | 5.14 |
| 3222 | 631 | 54862 | 174.56242 | 27.68206 | G2 | 16.88 | 5.76 |
| 3221 | 491 | 54864 | 175.42003 | 26.71263 | G0 | 16.79 | 5.06 |
| 3246 | 227 | 54939 | 176.13553 | 5.97447 | F9 | 17.11 | 5.53 |
| 3246 | 247 | 54939 | 176.24115 | 5.18235 | F5 | 18.14 | 5.24 |
| 3246 | 170 | 54939 | 176.62262 | 6.27239 | F5 | 16.59 | 5.01 |
| 3173 | 434 | 54849 | 177.60331 | 50.03237 | K5 | 18.43 | 5.27 |
| 3173 | 412 | 54849 | 178.07779 | 50.98405 | K1 | 17.35 | 5.6 |
| 3173 | 570 | 54849 | 179.23215 | 51.02912 | K3 | 17.55 | 5.02 |
| 3214 | 349 | 54866 | 180.52545 | 11.95398 | F9 | 16.18 | 5.12 |
| 3181 | 452 | 54860 | 180.72274 | 31.57088 | G2 | 16.82 | 5.14 |
| 3214 | 248 | 54866 | 181.20157 | 10.12625 | F5 | 17.31 | 5.24 |
| 3213 | 369 | 54865 | 181.20622 | 14.8449 | G2 | 16.16 | 5.23 |
| 3213 | 144 | 54865 | 181.50682 | 13.70937 | A0 | 19.47 | 5.62 |
| 3181 | 12 | 54860 | 181.79107 | 29.71221 | F9 | 15.92 | 5.06 |
| 3180 | 209 | 54864 | 182.47242 | 28.8078 | F9 | 17.01 | 5.3 |
| 3305 | 313 | 54945 | 183.07048 | 54.17639 | F9 | 16.85 | 5.42 |
| 3213 | 630 | 54865 | 183.10899 | 14.1723 | G2 | 16.45 | 5.05 |
| 3180 | 564 | 54864 | 184.03115 | 30.05302 | M5 | 21.71 | 5.3 |
| 3172 | 469 | 54863 | 184.09651 | 20.62191 | G0 | 16.71 | 5.19 |
| 3172 | 50 | 54863 | 184.67371 | 18.62552 | F2 | 16.25 | 5.04 |
| 3238 | 511 | 54885 | 185.58104 | 41.24556 | F9 | 16.77 | 5.11 |
| 3238 | 626 | 54885 | 186.35966 | 41.15343 | F5 | 17.07 | 5.44 |
| 3304 | 575 | 54942 | 186.83762 | 55.58244 | F9 | 17.54 | 5.79 |
| 3254 | 272 | 54889 | 187.64059 | 14.38726 | F9 | 16.34 | 5.52 |
| 3254 | 388 | 54889 | 187.72969 | 15.19258 | K1 | 18.11 | 5.27 |
| 3254 | 189 | 54889 | 188.188 | 14.09005 | G0 | 16.37 | 5.01 |
| 3254 | 572 | 54889 | 189.42706 | 15.47339 | F5 | 16.63 | 5.26 |
| 3367 | 371 | 54998 | 193.20184 | 48.1014 | F9 | 17.03 | 5.32 |
| 3234 | 246 | 54885 | 194.59811 | 8.56053 | F9 | 18.06 | 5.41 |
| 3236 | 552 | 54892 | 194.60564 | 30.38645 | F9 | 16.68 | 5.04 |
| 3237 | 413 | 54883 | 196.52464 | 60.53156 | F9 | 16.64 | 5.21 |
| 3303 | 51 | 54950 | 200.66292 | 24.13347 | G2 | 16.84 | 5.17 |
| 3318 | 391 | 54951 | 201.16319 | 54.26796 | G2 | 15.93 | 5.05 |
| 3318 | 234 | 54951 | 201.73013 | 53.89357 | F9 | 17.42 | 5.35 |
| 3318 | 420 | 54951 | 202.50007 | 54.76077 | F9 | 18.07 | 5.03 |
| 3318 | 168 | 54951 | 202.63017 | 53.05689 | F9 | 17.08 | 5.58 |
| 3318 | 486 | 54951 | 203.1801 | 54.1776 | F9 | 17.19 | 5,00 |
| 3318 | 535 | 54951 | 203.49582 | 54.56455 | K5 | 18.42 | 5.17 |
| 3003 | 282 | 54845 | 204.38884 | 47.08516 | K1 | 13.08 | 5.04 |
| 3318 | 550 | 54951 | 204.42066 | 54.1488 | G2 | 16.19 | 5.2 |
| 3310 | 69 | 54919 | 210.38372 | 17.75831 | G2 | **0,00** | 5.04 |
| 3312 | 250 | 54969 | 213.03633 | 45.28615 | K3 | 16.03 | 5.27 |
| 3313 | 449 | 54999 | 215.04127 | 10.15395 | G2 | 16.72 | 5.04 |
| 3296 | 433 | 54909 | 218.4031 | 56.38403 | F9 | 17.96 | 5.02 |
| 3296 | 73  | 54909 | 219.19144 | 53.98254  | F2 | 18.12 | 5.06 |
| 3296 | 71  | 54909 | 219.308   | 54.36972  | G2 | 16.02 | 5.03 |
| 3296 | 555 | 54909 | 219.48668 | 55.93023  | F9 | 18.39 | 5.71 |
| 3315 | 131 | 54942 | 223.84456 | 32.08184  | F2 | 15.85 | 5.11 |
| 3297 | 247 | 54941 | 227.86303 | 49.96438  | G2 | 17.49 | 5.11 |
| 3297 | 151 | 54941 | 229.39679 | 50.59978  | F2 | 15.87 | 5.2  |
| 3005 | 270 | 54876 | 239.70256 | 27.46743  | G2 | 11.24 | 5.03 |
| 3005 | 162 | 54876 | 240.38894 | 27.56526  | F9 | 12.55 | 5.36 |
| 3291 | 291 | 54939 | 252.06355 | 40.95487  | K1 | 17.04 | 5.27 |
| 3291 | 226 | 54939 | 252.4945  | 41.05637  | F9 | 17.26 | 5.09 |
| 3291 | 62  | 54939 | 254.57143 | 41.16952  | G2 | 18.1  | 5.54 |
| 3143 | 45  | 54772 | 343.67051 | -0.96321  | G2 | 15.84 | 5.14 |
| 3142 | 447 | 54735 | 353.81723 | 1.01613   | F5 | 17.57 | 5.08 |
| 3144 | 297 | 54763 | 354.0782  | -10.95415 | F9 | 17.22 | 5.53 |
| 3130 | 243 | 54740 | 354.94178 | 13.86028  | F9 | 17.56 | 5.22 |
| 3144 | 599 | 54763 | 356.05048 | -9.47781  | F9 | 16.51 | 5.21 |
| 3130 | 518 | 54740 | 356.10602 | 15.2327   | K1 | 17.71 | 5.27 |
| 3144 | 625 | 54763 | 356.38786 | -9.56591  | F9 | 16.02 | 5.02 |
| 3131 | 450 | 54731 | 357.78126 | 15.0766   | F9 | 18.6  | 5.21 |